\documentclass[sigconf, nonacm]{acmart}

\acmYear{2026}
\usepackage{xspace}
\usepackage{listings, algorithmic}
\usepackage[framemethod=TikZ]{mdframed}
\usepackage{enumitem}
\usepackage{relsize}
\usepackage{xargs}
\usepackage[most]{tcolorbox}
\usepackage{multirow}
\usepackage{xfrac}
\usepackage{makecell}
\usepackage{amsthm}
\usepackage{cases}
\usepackage[official]{eurosym}
\usepackage{makecell}
\usepackage{pifont}

\newcommand{\xmark}{\ding{55}}  % ✘
\newcommand{\circled}[1]{\raisebox{.5pt}{\textcircled{\raisebox{-.9pt} {#1}}}}

\usepackage{tikz} \usetikzlibrary{positioning,shapes.geometric}

\newcommand{\comments}[1]{\textcolor{gray}{\texttt{#1}}}
\newcommand{\shortrightarrow}[1][3pt]{\hspace{2pt}\mathrel{%
    \hbox{\rule[\dimexpr\fontdimen22\textfont2-.2pt\relax]{#1}{.4pt}}%
\mkern-4mu\hbox{\usefont{U}{lasy}{m}{n}\symbol{41}}}}

\newtheorem{theorem}{Theorem}

\theoremstyle{definition}
\newtheorem{definition}{Definition}[section]

\newcounter{protocol}
\newenvironment{protocol}[2]{%
  \refstepcounter{protocol}\label{#2}
  \mdfsetup{%
    frametitle={%
      \tikz[baseline=(current bounding box.east),outer sep=0pt]
      \node[anchor=east,rectangle,fill=black!0]
    {\strut Protocol~\theprotocol:~#1};}%
  }
  \mdfsetup{%
    innertopmargin=0pt,linecolor=black,%
    linewidth=1pt,topline=true,%
    frametitleaboveskip=\dimexpr-\ht\strutbox\relax%
  }
  \begin{figure*}
  \begin{mdframed}[]\relax}{%
    \end{mdframed}
\end{figure*}}

\newenvironment{protocolOne}[2]{%
  \refstepcounter{protocol}%
  \label{#2}%
  \mdfsetup{%
    frametitle={%
      \tikz[baseline=(current bounding box.east),outer sep=0pt]
        \node[anchor=east,rectangle,fill=black!0]
        {\strut Protocol~\theprotocol:~#1};}%
  }%
  \mdfsetup{%
    innertopmargin=0pt,linecolor=black,%
    linewidth=1pt,topline=true,%
    frametitleaboveskip=\dimexpr-\ht\strutbox\relax%
  }%
  \begin{figure}[t]
  \begin{mdframed}[]\relax
}{%
  \end{mdframed}
  \end{figure}%
}

\newcounter{simulation}
\newenvironment{simulation}[2]{%
  \refstepcounter{simulation}\label{#2}
  \mdfsetup{%
    frametitle={%
      \tikz[baseline=(current bounding box.east),outer sep=0pt]
      \node[anchor=east,rectangle,fill=black!0]
    {\strut Simulator~\thesimulation:~#1};}%
  }
  \mdfsetup{%
    innertopmargin=0pt,linecolor=black,%
    linewidth=1pt,topline=true,%
    frametitleaboveskip=\dimexpr-\ht\strutbox\relax%
  }
   \begin{figure}[tbhp]
  \begin{mdframed}[]\relax}{%
    \end{mdframed}
 \end{figure}}

\newcounter{functionality}
\newenvironment{functionality}[2]{%
  \global\refstepcounter{functionality}\label{#2}
  \mdfsetup{%
    frametitle={%
      \tikz[baseline=(current bounding box.east),outer sep=0pt]
      \node[anchor=east,rectangle,fill=black!0]
    {\strut Functionality~\thefunctionality:~#1};}%
  }
  \mdfsetup{%
    innertopmargin=0pt,linecolor=black,%
    linewidth=1pt,topline=true,%
    frametitleaboveskip=\dimexpr-\ht\strutbox\relax%
  }
  \begin{figure*}
  \begin{mdframed}[]
  \relax}{%
    \end{mdframed}
\end{figure*}}

\newenvironment{functionalityOne}[2]{%
  \refstepcounter{functionality}\label{#2}
  \mdfsetup{%
    frametitle={%
      \tikz[baseline=(current bounding box.east),outer sep=0pt]
      \node[anchor=east,rectangle,fill=black!0]
    {\strut Functionality~\thefunctionality:~#1};}%
  }
  \mdfsetup{%
    innertopmargin=0pt,linecolor=black,%
    linewidth=1pt,topline=true,%
    frametitleaboveskip=\dimexpr-\ht\strutbox\relax%
  }
  \begin{figure}
  \begin{mdframed}[]
  \relax}{%
    \end{mdframed}
\end{figure}
}

\NewDocumentEnvironment{example}{ O{} }
{\colorlet{colexam}{red!55!black}
  \newtcolorbox[use counter=example]{examplebox}{%
    empty,% Empty previously set parameters
    title={Example: #1},% use \thetcbcounter to access the testexample counter text
    attach boxed title to top left,
    minipage boxed title,
    boxed title style={empty,size=minimal,toprule=0pt,top=4pt,left=3mm,overlay={}},
    coltitle=colexam,fonttitle=\bfseries,
    before=\par\medskip\noindent,parbox=false,boxsep=0pt,left=3mm,right=0mm,top=2pt,breakable,pad at break=0mm,
    before upper=\csname @totalleftmargin\endcsname0pt, % Use instead of parbox=true. This ensures parskip is inherited by box.
    overlay unbroken={\draw[colexam,line width=.5pt] ([xshift=-0pt]title.north west) -- ([xshift=-0pt]frame.south west); },
    overlay first={\draw[colexam,line width=.5pt] ([xshift=-0pt]title.north west) -- ([xshift=-0pt]frame.south west); },
    overlay middle={\draw[colexam,line width=.5pt] ([xshift=-0pt]frame.north west) -- ([xshift=-0pt]frame.south west); },
    overlay last={\draw[colexam,line width=.5pt] ([xshift=-0pt]frame.north west) -- ([xshift=-0pt]frame.south west); },%
  }
\begin{examplebox}}{%
\end{examplebox}\endlist}

\usepackage[colorinlistoftodos,prependcaption,textsize=tiny,textwidth=0.8in]{todonotes} %,disable

\newcommandx{\unsure}[2][1=]{\todo[linecolor=red,backgroundcolor=red!25,bordercolor=red,#1]{#2}}
\newcommandx{\change}[2][1=]{\todo[linecolor=blue,backgroundcolor=blue!25,bordercolor=blue,#1]{#2}}
\newcommandx{\info}[2][1=]{\todo[linecolor=green,backgroundcolor=green!25,bordercolor=green,#1]{#2}}
\newcommandx{\improvement}[2][1=]{\todo[linecolor=purple,backgroundcolor=purple!25,bordercolor=purple,#1]{#2}}
\newcommandx{\bart}[2][1=]{\todo[linecolor=orange,backgroundcolor=orange!25,bordercolor=red,#1]{Bart: #2}}
\newcommandx{\krive}[2][1=]{\todo[linecolor=green,backgroundcolor=green!25,bordercolor=black,#1]{Krive: #2}}
\newcommandx{\jl}[2][1=]{\todo[linecolor=blue,backgroundcolor=blue!25,bordercolor=black,#1]{JL: #2}}
\newcommandx{\fk}[2][1=]{\todo[linecolor=yellow,backgroundcolor=yellow!25,bordercolor=orange,#1]{FK: #2}}
\newcommandx{\cdsg}[1]{\todo[inline,linecolor=olive,backgroundcolor=olive!25,bordercolor=black]{CD: #1}}

\usepackage{tcolorbox}
\newtcolorbox{remarkbox}{colback=yellow!20!white, colframe=red!50!black,
  boxrule=1pt, arc=4pt, auto outer arc, fontupper=\small,
  breakable, left=2pt, right=2pt, top=2pt, bottom=2pt}

\newcommand{\party}[1]{\ensuremath{P^{#1}}\xspace}
\newcommand{\parties}[0]{\ensuremath{\mathbf P}\xspace}
\newcommand{\Pcor}[0]{\ensuremath{{\mathbf P^*}}\xspace}

\newcommand{\thirdparty}[0]{\ensuremath{\mathit{C}}\xspace}

\newcommand{\List}{\mathcal{L}}

\newcommand{\collector}[0]{\ensuremath{\mathit{C}}\xspace}
\newcommand{\IntersectionInfoSet}{{Y}_{\cap}}
\newcommand{\TPInputSet}{{R}}
\newcommand{\TPSimulator}{\simulator_{\thirdparty}}

\newcommand{\distinguisher}[0]{\ensuremath{D}\xspace}
\newcommand{\simulator}[0]{\ensuremath{Sim}\xspace}
\newcommand{\adversary}[0]{\ensuremath{\mathcal{A}}\xspace}
\newcommand{\environment}[0]{\ensuremath{\mathcal{Z}}\xspace}
\newcommand{\set}[1]{\ensuremath{E_{#1}}\xspace}
\newcommand{\vect}[0]{\ensuremath{{V}}\xspace}
\newcommand{\Perm}[0]{\ensuremath{\mathlarger\pi}}

\newcommand{\enc}{\ensuremath{\mathsf{enc}}\xspace}

\newcommand{\dec}[0]{\ensuremath{\mathsf{dec}}\xspace}
\newcommand{\prp}[1]{\ensuremath{\mathsf{prp_{#1}}}\xspace}
\newcommand{\iprp}[1]{\ensuremath{\mathsf{prp^{-1}_{#1}}}\xspace}
\newcommand{\negl}[0]{\ensuremath{\mathsf{negl}}\xspace}
\newcommand{\tssrecon}[1]{\ensuremath{\mathsf{recon_{#1}}}\xspace}
\newcommand{\tsssplit}[1]{\ensuremath{\mathsf{split_{#1}}}\xspace}
\newcommand{\encode}[1]{\ensuremath{\mathsf{encode}_{#1}}\xspace}
\newcommand{\decode}[1]{\ensuremath{\mathsf{decode}_{#1}}\xspace}

\newcommand{\Fcore}{\ensuremath{\mathcal{F}_{SIKA}}\xspace}
\newcommand{\Pcore}{\ensuremath{\Pi_{SIKA}}\xspace}
\newcommand{\Fpay}{\ensuremath{\mathcal{F}_{payload}}\xspace}
\newcommand{\Ppay}{\ensuremath{\Pi_{payload}}\xspace}

\newcommand{\ID}[1]{\ensuremath{\mathit{X}_{#1}}\xspace}
\newcommand{\records}[1]{\ensuremath{{Recs_{#1}}}\xspace}
\newcommand{\Atts}[2]{\ensuremath{\mathit{A{tt}s}^{#1}_{#2}}\xspace}

\newcommand{\id}[2]{\ensuremath{\mathit{id}^{#1}_{#2}}\xspace}

\newcommand{\key}[1]{\ensuremath{\mathit{key}^{#1}}\xspace}

\newcommand{\share}[2]{\ensuremath{\mathit{s}^{#1}_{#2}}\xspace}
\newcommand{\bshare}[3]{\ensuremath{\mathit{bs}^{#1\shortrightarrow#2}_{#3}}\xspace}
\newcommand{\boshare}[3]{\ensuremath{\widetilde{\mathit{bs}}^{#1\shortrightarrow#2}_{#3}}\xspace}

\newcommand{\bsecret}[2]{\ensuremath{\mathit{bnym}^{#1}_{#2}}\xspace}
\newcommand{\bosecret}[2]{\ensuremath{\mathit{b}\widetilde{\mathit{nym}}^{#1}_{#2}}\xspace}

\newcommand{\secret}[1]{\ensuremath{\mathit{nym}_{#1}}\xspace}
\newcommand{\secrets}[1]{\ensuremath{\mathit{Nym}^{#1}}\xspace}
\newcommand{\osecret}[2]{\ensuremath{\widetilde{\mathit{nym}}^{#1}_{#2}}\xspace}

\newcommand{\rnd}[3]{\ensuremath{\mathit{z}^{#1\shortrightarrow#2}_{#3}}\xspace}
\newcommand{\Rnd}[3]{\ensuremath{\mathit{\bar Z}^{#1\shortrightarrow#2}_{#3}}\xspace}
\newcommand{\Rndm}[1]{\ensuremath{\mathit{\bar Z}_{#1}}\xspace}
\newcommand{\ornd}[3]{\ensuremath{\Tilde{\mathit{z}}^{#1\shortrightarrow#2}_{#3}}\xspace}
\newcommand{\oRnd}[3]{\ensuremath{\Tilde{\bar Z}^{#1\shortrightarrow#2}_{#3}}\xspace}
\newcommand{\shsecret}[2]{\ensuremath{\mathit{v}^{#1}_{#2}}\xspace}
\newcommand{\shshare}[2]{\ensuremath{\mathit{sh}^{#1}_{#2}}\xspace}

\newcommand{\inp}[2]{\ensuremath{\mathit{input}_{#2}^{#1}}\xspace}
\newcommand{\tape}[2]{\ensuremath{\mathit{tape}_{#2}^{#1}}\xspace}
\newcommand{\msg}[2]{\ensuremath{\mathit{msg}_{#2}^{#1}}\xspace}
\newcommand{\outp}[2]{\ensuremath{\mathit{output}_{#2}^{#1}}\xspace}
\newcommand{\view}[2]{\ensuremath{\mathit{view}_{#2}^{#1}}\xspace}

\newcommand{\result}{\ensuremath{\mathit{R}}\xspace}

\newcommand{\okvs}[2]{\ensuremath{\mathit{okvs}^{#1\shortrightarrow#2}\xspace}}

\begin{document}

\title[Labeled Delegated PSI]{Labeled Delegated PSI and its Applications in the Public Sector}
\author{Kristof Verslype}
\affiliation{%
\institution{Smals Research}
\city{Brussels}
\country{Belgium}
}
\email{kristof.verslype@smals.be}
\author{Florian Kerschbaum}
\affiliation{%
\institution{University of Waterloo}
\city{Waterloo}
\country{Canada}
}
\email{florian.kerschbaum@uwaterloo.ca}
\author{Cyprien Delpech de Saint Guilhem}
\affiliation{%
\institution{Cosic - KU Leuven, 3MI Labs}
\city{Leuven}
\country{Belgium}
}
\email{cyprien.delpechdesaintguilhem@kuleuven.be}
\author{Bart De Decker}
\affiliation{%
\institution{DistriNet - KU Leuven}
\city{Leuven}
\country{Belgium}
}
\email{bart.dedecker@cs.kuleuven.be}
\author{Jorn Lapon}
\affiliation{%
\institution{DistriNet - KU Leuven}
\city{Gent}
\country{Belgium}
}
\email{jorn.lapon@kuleuven.be}

\renewcommand{\shortauthors}{Verslype et al.}

\begin{abstract}
  Sensitive citizen data, such as social, medical, and fiscal data, is heavily fragmented across
  public bodies and the private domain. Mining the combined data sets allows for new insights that otherwise remain hidden.
  Examples are improved healthcare, fraud detection, and evidence-based policy making.
  (Multi-party) delegated private set intersection (D-PSI) is a privacy-enhancing technology to link data across multiple data providers using a data collector.
  However, before it can be deployed in these use cases, it needs to be enhanced with additional functions, e.g., securely delivering payload only for elements in the intersection.
  Although there has been recent progress in the communication and computation requirements of D-PSI, these practical obstacles have not yet been addressed.
  
  This paper is the result of a collaboration with a governmental organization responsible for collecting, linking, and pseudonymizing data.
  Based on their requirements, we design a new D-PSI protocol with composable output functions, including encrypted payload and pseudonymized identifiers.
  We show that our protocol is secure in the standard model against colluding semi-honest {\em data providers} and against a non-colluding, possibly malicious independent party, the {\em data collector}.
  It, hence, allows to privately link {\em and collect} data from multiple data providers suitable for deployment in these use cases in the public sector.
\end{abstract}


\maketitle

\section{Introduction}
\label{sct:introduction}
Data is very often fragmented across government agencies and commercial entities.
However, linking this data is usually challenging, as the data may contain sensitive, personally identifiable information (PII), 
% and privacy regulations should be respected. BDD
making it extra difficult to meticulously respect the privacy regulations.

As an example, an academic researcher investigates whether citizens, with a yearly salary above \euro{}50.000, who are self-employed as a secondary occupation and were born in 1990 or later, are sufficiently insured in their capacity of a self-employed person against certain occupational diseases.
Therefore, the researcher is interested in non-aggregated -- i.e. non-summarized -- pseudonymized data about this subgroup of people.
However, fiscal, employment, and insurance data are managed by separate organizations (data providers).
Hence, this basic query already involves at least three data providers, that are unable to select the relevant records on their own. 

In practice, for privacy reasons, existing solutions turn to a combination of trusted parties and strong legal regulations about the collection, linking and pseudonymization of the data, before data can be used for analysis.
Examples of parties trusted to perform data linking are the Census Bureau in the US, the Research Data Centers (RDCs) in Germany, the National Institute of Statistics and Economic Studies (INSEE) in France and the Federal Statistical Office in Switzerland \cite{FORS}.
However, these trusted parties are themselves interested in reducing their risks and aim to use privacy-enhancing technologies (PETs).

Private set intersection (PSI) emerges as a promising privacy-enhancing technology to address these challenges effectively.
Mora\-les et al.~\cite{escalera23} present an excellent survey of PSI protocols, distinguishing between two-party PSI, multi-party PSI (M-PSI) and delegated (or outsourced) PSI (D-PSI).
Two-party PSI is the common case in the literature where two parties privately compute the intersection of their sets of identifiers, while revealing nothing about the identifiers that do not belong to the intersection.
M-PSI~\cite{kissner2004private,kissner2005privacy,hazay2017scalable,inbar2018efficient,kavousi2021efficient,ben2022psimple,zhang2019efficient,vos2022fast,kolesnikov2017practical,wang2021multi,bay2021practical,mahdavi2020practical,badrinarayanan2021multi,CDGOSS21} 
allows multiple (more than two) parties to do so.
However, many existing M-PSI protocols are not suitable for deployment by the entities that are entrusted to collect such data while not having any input data of their own.
Instead, D-PSI \cite{delegatable,Kerschbaum12a,Kerschbaum12b,Dong13,Abadi15,Abadi16,Abadi19} -- or variants thereof -- seem suitable for this case where the intersection is computed at a third party having no input.
In particular, D-PSI protocols where multiple (more than two) parties can submit their inputs are relevant in this use case.

Yet, the intersection of identifiers is insufficient information for this use case.
The entrusted entity requires additional information about records associated with the identifiers.
In two-party PSI, this can be accomplished by labeled PSI \cite{Bienstock24,cong2021labeled,chen2018labeled} where a label is revealed for each identifier in the intersection.
This label can be a cryptographic key used to encrypt the additional information in the records to be revealed if the element is in the intersection.
However, labeled PSI does not exist for D-PSI or M-PSI protocols.
Circuit-PSI, a variant that allows computing any function over the intersection, including revealing cryptographic keys, exists in the M-PSI case~\cite{CDGOSS21}, but not in the D-PSI case.
Furthermore, revealing labels using circuit-PSI is less efficient than using dedicated labeled PSI protocols.

To understand why labeled PSI is non-trivial in the D-PSI case, consider the following complications of the common construction in the two-party case.
To encrypt its information a party must have access to the key.
In two-party PSI based on oblivious key-value stores (OKVS) -- or variants, such as oblivious programmable pseudo-random functions (OPPRF) -- this is trivial.
The sender, i.e., the holder of the key-value stores, chooses the key and uses it to encrypt its data.
The recipient obtains the key if and only if it has the matching identifier.
In this case, there is only one key (per identifier), one sender, and one recipient.
However, in D-PSI, there are multiple senders and no two sending parties should have the same key, since revealing that key would compromise the confidentiality of the encrypted information.
Any key should only be revealed and only to the entrusted party if \underline{all}\ parties have the corresponding identifier, i.e., a subset of parties with the same identifier should not reveal any keys.
Hence, there is no single other party with whom to compute a ``partial intersection'', but instead each party needs to intersect with all other parties, and there is no straightforward reduction from the multi-party case to the two-party case.
\paragraph{Contributions.}
%\remark{FK: This lower part can still be improved.}
First, to solve these challenges, we design and prove a new multi-party {\em Set Intersection Key Agreement} (SIKA) protocol. Second, we show how to use this protocol as a composable functionality when building several practical output protocols, including Labeled D-PSI. 
Finally, the protocol has been implemented and applied to our use case, \textsf{LetheLink}, demonstrating the practicality of the solution.

The outline of the paper is as follows: we first provide a practical context and requirements in Section~\ref{sec:req}, followed by related work and general information in Section~\ref{sct:related} and Section~\ref{sct:preliminaries} respectively. Section~\ref{sct:sika} presents the SIKA protocol and its security proof, followed by output protocols in Section~\ref{sct:output}. Finally, \textsf{LetheLink} is presented and evaluated in Section~\ref{sct:lethelink} followed by conclusions in Section~\ref{sct:conclusion}.

\section{Practical Context and Requirements}\label{sec:req}
In current practice, for privacy reasons, linking data from different (often public) organizations is resolved through a combination of a trusted party, and strong legal regulations. In a next step, the trusted party offers a controlled release of the linked data.
Due to regulations such as GDPR and HIPAA, trusted parties aim to use PETs to lower their risks, and mitigate the inherent dangers of managing highly sensitive and valuable information.

This paper stems from a collaboration with a governmental institution tasked with collecting, linking and pseudonymizing citizen, tax and health data; with the primary objective of making D-PSI deployable for their practice.

\noindent We first briefly present current practices, followed by the requirements of an enhanced privacy friendly alternative.

\paragraph{Current Practices.}
Our partner's current methodology consists of four distinct steps.

{\bf Step 1:} A researcher needs to access the combined data, linked from multiple data providers, for a particular research project. To that end, a complex query is proposed that will select and combine data from different data providers. The purpose of the research project is documented and the query justified and submitted to the scientific board for approval.

{\bf Step 2:} The scientific board investigates the requests from researchers and (dis)approves them. The board consists of scientists, lawyers, privacy specialists and possibly a delegation of the data providers involved. It may impose restrictions on the data collection and even prescribe additional privacy-enhancing post-processing to be performed by the trusted party before the data is presented to the researcher.

{\bf Step 3:} The approval is sent to the data providers (i.e., input parties) who prepare their data (filter attributes and records that can be filtered locally). The data providers send their data including identifiers to the trusted party. The trusted party is in charge of performing the linking, pruning and removing the identifiers.

{\bf Step 4:} The trusted party cleans and processes the data before presenting the data to the researcher. Depending on the privacy requirements, the researcher may only access the data in a secure and trusted environment or the data is only made available through differential privacy solutions~\cite{D08}.

This approach requires high trust in the trusted party, since it has access to much more personal data than the final result. Moreover, unwanted data leakage may potentially occur to the individual data providers depending on how the data is retrieved from those input parties.\\

\paragraph{D-PSI as a privacy enhancing alternative.}
PSI is very diverse and research has led to many different solutions for different problems. A delegated, multi-party variant, D-PSI, is a promising candidate for this use case. It addresses a number of important privacy requirements and current constructions offer practical performance metrics. Current D-PSI constructions, however, lack either all or some of the features that makes them a true fit for our trusted party scenario.
\\
We list the requirements below:

\paragraph{Minimal, but necessary requirements.}
\begin{enumerate}[label=R\arabic*]
    \item {\bf Data Collector Only.} \label{req:tp} None of the participating data providers (P) should gain any new information, such as the size or characteristics of another party's database. Thus, it may not learn whether an identifier belongs to a partial or the full intersection; it may not even learn the sizes of these intersections. The result is only received by a (semi-trusted; i.e., non-colluding) data collector \collector.

    \item {\bf Local.} Because of privacy regulations, data providers do not share their database with external parties (except for what belongs to the intersection with the data collector).
    
    \item {\bf Pseudonymous.}\label{req:nym} No real identifiers are revealed to the data collector \thirdparty. The resulting pseudonyms enable communication between data providers and the data collector regarding specific records.
    
    \item {\bf Linking.}\label{req:link} The data collector is the only party in the protocol that can link records. In current practice, the party receiving the results is responsible for removing the identifiers after linking records. Currently, most M-PSI solutions result in a set of identifiers in the intersection being revealed to one of the parties.

    \item {\bf Private Payload Transfer.}\label{req:payload} For each identifier \id {} k in the intersection, the data collector learns the corresponding record \(Atts_k^i\) in clear text held by each data provider \party i and, hence, may link them. When not in the intersection, the payload is not disclosed.

    \item {\bf Scalable.} The protocol should scale well, both computationally and what concerns network communication. It should be able to handle large databases with at least a million records and several data providers.

\end{enumerate}
\paragraph{Optional, but highly desired requirements.}
\begin{enumerate}[label=R\arabic*,resume]
  \item {\bf Intersection Threshold.} It should be possible to specify a minimal cardinality of the intersection before the payload will be disclosed. Due to varying levels of trust and data sensitivity, this threshold may be different for different data providers.
  \item {\bf Re-Identification.} In some cases, under strong legal requirements, the protocol should allow for re-identification (e.g., for correcting errors, for law enforcement, etc.).
  
\end{enumerate}

While the properties of traditional D-PSI may not be sufficient, many works have implemented some of these requirements.
Resolving the data collector-only requirement (\ref{req:tp}), a D-PSI protocol not only needs to delegate the computation of the intersection, but should also allow decryption of the intersection payload at the data collector.  This is an extension of the generic D-PSI protocol as identified by Morales et al.~\cite{escalera23}.

%one-way M-PSI, has only one receiving party. However, to fully meet \ref{req:tp}, the receiving party would have to provide the universe \(\mathcal{U}\) as its input set which could be efficient with garbled bloom filters with an all-one filter.

%Other solutions for \ref{req:tp} are \emph{Third Party} PSI~\cite{YY23} using public key cryptography and M-PSI with \emph{External Decider}~\cite{RMNBG21} for use cases with a universe of only limited size.
%While these approaches address aspects of the third-party requirement (\ref{req:tp}), they fall short in meeting the broader set of needs outlined above, particularly in scalability and support for pseudonymity.

%Alternatively, input parties may use circuit M-PSI and send a function of the intersection to the third party such that only this party may obtain the intersection.
%While this is a possibility, meeting \ref{req:nym} (pseudonymous) and \ref{req:link} (linking) requires substantial overhead and complex processing, making it impractical for large-scale applications.

Labeled PSI offers a promising direction to address our requirements for \ref{req:nym} and \ref{req:link}.
Considering \ref{req:nym} and \ref{req:link}, in labeled PSI, the receiver obtains a label for each identifier provided by the data source.
This label could be used as a cryptographic pseudonym to enable secure payload transfer (\ref{req:payload}), for instance, by serving as decryption keys. 

However, to the best of our knowledge, a D-PSI variant of labeled PSI, and in fact, a clear definition of what labeled PSI in the multi-party settings entails, does not exist. Assuming in labeled D-PSI each data source decides upon its own labels, and the receiving party may link them together for the same identifier, an efficient implementation is not straightforward, nor does it support \ref{req:nym}, in which the identifiers are kept undisclosed.

There is a vast amount of efficient D-PSI (and M-PSI) solutions (see also Related Work in Section~\ref{sct:related}). While these solutions either include some of these requirements or may somehow be extended to include them, {\it there is no straightforward solution that covers for all of these requirements}.
These gaps motivate our development of a novel, practical protocol designed to comprehensively meet the outlined requirements, enabling secure and efficient deployment in real-world scenarios.

\section{Related Work}\label{sct:related}

Private Set Intersection (PSI) \cite{meadows,freedman,vole,raghuraman2022blazing} allows two parties to compute the intersection of their data sets without revealing any item not in the intersection.
This problem has been studied since the 1980s \cite{meadows} and has been adopted in industrial applications \cite{psigoogle15}.
The term was popularized by Freedman et al.~\cite{freedman}.
There has been vast research on improving the efficiency and functionality of PSI.
The currently fastest protocols use oblivious pseudo-random functions (OPRF) constructed from vector-oblivious linear evaluations (VOLE) \cite{vole,raghuraman2022blazing}.
Morales et al.~\cite{escalera23} present an excellent survey of PSI protocols, distinguishing between two-party PSI, multi-party PSI (M-PSI) and delegated (or outsourced) PSI (D-PSI).

Two-party PSI is the common case in the literature where two parties privately compute the intersection of their sets of identifiers, but nothing about identifiers not in the intersection is revealed.
% In M-PSI~\cite{kissner2004private,kissner2005privacy}, where the problem was first introduced, various scalable and efficient protocols have since ben proposed~\cite{hazay2017scalable,inbar2018efficient,kavousi2021efficient,ben2022psimple,zhang2019efficient,vos2022fast,kolesnikov2017practical,wang2021multi,bay2021practical,mahdavi2020practical,badrinarayanan2021multi,CDGOSS21}, more than two parties maintain a data set and want to compute their intersection. This problem was first studied by Kissner and Song~\cite{kissner2004private,kissner2005privacy}. 
The multi-party private set intersection (M-PSI) problem, first introduced by Kissner and Song~\cite{kissner2004private,kissner2005privacy}, involves more than two parties —each holding their own dataset— who wish to compute the intersection of their sets. Since then, several scalable and efficient protocols have been proposed~\cite{hazay2017scalable,inbar2018efficient,kavousi2021efficient,ben2022psimple,zhang2019efficient,vos2022fast,kolesnikov2017practical,wang2021multi,bay2021practical,mahdavi2020practical,badrinarayanan2021multi,CDGOSS21}.
Scalability and privacy become more complex in M-PSI.

Specialized D-PSI protocols have been developed where the intersection is computed at a third party.
Kerschbaum introduced D-PSI as outsourced PSI~\cite{Kerschbaum12b}.
He presented a variant where the third party obtains the intersection and one where the third party is blind to the intersection.
Later, he also presented using additively homomorphic encryption~\cite{Kerschbaum12a}.
Dong et al.~\cite{Dong13} later reintroduced the same concept and called the third party a semi-trusted arbiter.

Abadi et al.~\cite{Abadi15,Abadi19} reintroduced the concept again and called it delegatable PSI (D-PSI), the name adopted in Morales et al.'s survey.
Note that Morales et al.~only consider the version where the third party is blind to the intersection, and while many protocols are also applicable in the other model introduced by Kerschbaum, a black-box reduction is not obvious.
Abadi et al.~updated the features of their protocols to include updates of the stored elements~\cite{delegatable} and verifiability of the intersection~\cite{Abadi16}.

\paragraph{Evaluation of Feature Support} In this paper, we outlined several requirements, many of which have been addressed individually in prior work, but are rarely supported in combination. While Section~\ref{sec:req} defines the full set of requirements, here we highlight a subset that are particularly relevant for comparing with related work. 
Specifically, we consider from Section~\ref{sec:req} for convenience: (i) support for multiple parties contributing input, (ii) that only the collector, not providing input, learns the protocol's outcome (R1), (iii) that actual identifiers remain hidden and are replaced with pseudonyms (R3), and that the protocol supports the transfer of payloads associated with each identifier in the intersection (R5).

While R1 could in principle be obtained from M-PSI via a black-box reduction to D-PSI with a collector by introducing a third party whose input are all elements in the domain, this is clearly inefficient and does not fulfill the other requirements.

Chandran et al.~\cite{CDGOSS21} present a protocol that offers an interesting variant: circuit M-PSI.
In circuit M-PSI, each party obtains a secret share of a bit indicating whether the element is in the intersection or not.
This bit can then be used to selectively generate output (including shares for a third party), such as the labels (payload) in our protocol.
Both two-party~\cite{vole,CGS22,pinkas_efficient_2019} and multi-party variants exist~\cite{CDGOSS21}.
However, outputting one bit has a communication complexity of $O(n^2)$ and we require a label per party ($n$) and element ($m$) totaling a communication complexity of $O(n^3 m)$.

Table~\ref{tab:rw} summarizes representative work on D-PSI and Labeled PSI and our core requirements R1, R3 and R5. We discuss the key categories below.\\
First, we observe two flavors of {\it two-party D-PSI} (\circled{1} and \circled{2}). While they propose a third party for the computation, this party does not obtain the result (R1). In fact, one of the main goals is that this party does not learn anything. Hung et al. present a solution that supports R3 and R5, as it implements circuit PSI output~\cite{le_two-party_2019}, but it leaks the intersection size to the third party. 

Similarly, in the {\it multi-party D-PSI} domain, lots of prior work lacks the required features (\circled{3}). Sharma et al, present PRISM~\cite{sharma_prism_2024}, a multi-party delegated circuit-PSI solution (\circled{4}). The data is secret shared across multiple non-colluding public servers. While R3 and R5 could be solved (inefficiently), the use of multiple public servers contradicts requirement R2, to prevent sharing of data.

Another variant of D-PSI, was introduced by Kerschbaum~\cite{Kerschbaum12b} where the third party, referred to as {\it the collector}, obtains the intersection. This approach is also known as PSI with an external decider~\cite{RMNBG21} or third party PSI~\cite{yeo_third-party_2023,yeo_round-optimal_2024,liang_practical_2024,ye_quantum_2024}. Both variants for two-party (\circled{5}) and multi-party PSI (\circled{6}) are available. Unfortunately, support for R3 and R5 is non-trivial and lacking.

Finally, for {\it labeled PSI}, a two-party protocol, there are currently no D-PSI versions in which a collector obtains the result (\circled{7}). They may support the payload requirement (R5), but only Karakoç et al.~\cite{karakoc_linear_2020} provide a circuit PSI variant that allows pseudonymization of the identifiers (\circled{8}). No multi-party variants exist.

\begin{table}[h!]
\caption{Comparison of Delegated and Labeled PSI Variants}\label{tab:rw}
% \centering
\begin{tabular}{ll|c|c|c|c}
\hline
&\textbf{Related Work} &\textbf{M-PSI} & \textbf{R1} & \textbf{R3} & \textbf{R5}  \\
\hline
\circled{1}
 &\cite{oliaiy_verifiable_2017,wang_tag-based_2022,yang_improved_2018,wang_privacy-preserving_2018, wang_faster_2018, ali_attribute-based_2020, wang_private_2021,Kerschbaum12a,Dong13}    & \xmark & \xmark & \xmark & \xmark\\
\circled{2}& \cite{le_two-party_2019}      & \xmark & \xmark & \checkmark & \checkmark  \\ %Circuit psi
\circled{3}  & \cite{terada_improved_2018,miyaji_privacy-preserving_2017,qiu_ppsi_2019,adavoudi_jolfaei_eo-psi-ca_2022,kamara_scaling_2014,ruan_efficient_2020,
zhang_server-aided_2017,chen_two_2019,Abadi19,delegatable,Abadi15,Abadi16}   & \checkmark & \xmark & \xmark & \xmark \\
% \circled{4} & \cite{li_outsourced_2019}     & \checkmark & \xmark  & \xmark& \checkmark  \\ Cardinality
\circled{4} &\cite{sharma_prism_2024}      & \checkmark & \xmark & \checkmark & \checkmark   \\ %$\Delta$
\circled{5} & \cite{yeo_third-party_2023,yeo_round-optimal_2024,liang_practical_2024,ye_quantum_2024}  & \xmark & \checkmark     & \xmark & \xmark  \\
\circled{6} & \cite{Kerschbaum12b,RMNBG21}    & \checkmark & \checkmark     & \xmark & \xmark \\\hline
\circled{7} & \cite{chen2018labeled,cong2021labeled,Bienstock24}      & \xmark     & \xmark & \xmark& \checkmark   \\
\circled{8} & \cite{karakoc_linear_2020}     & \xmark     & \xmark & \checkmark & \checkmark  \\
\hline
\end{tabular}\caption*{{\small R1: Collector without input;\\
R3: pseudonymous intersection;\\
R5: payload support.}}
\end{table}
To our knowledge, this is the first practical protocol to meet R1, R3 and R5 simultaneously in a delegated multi-party setting.
In this paper, we investigate the functional extensions required for deploying D-PSI where a collector obtains the linked data without revealing the identifier.
We design a new protocol using symmetric key cryptography, implementing our requirements set forth and demonstrate practical performance results.\\

In our protocol, we link data sources using a common identical identifier.
While our focus is on exact matches, real-world deployments sometimes require approximate matching on joint attributes, such as name, birthday, telephone number, etc.
Protocols for approximate matching are known as private record linkage, e.g., \cite{WeiKer23,KhuKer20,HeMFS17,StaKSSTKHL22}, or fuzzy-PSI, e.g., \cite{UzuCKBL21,ChaFR23}.
However, their communication and computation complexity is significantly higher, since more elements need to be compared for matches.
\section{Preliminaries}\label{sct:preliminaries}
\noindent This Section sets the general notation, explains the required cryptographic primitives and elaborates on the security model.

\paragraph{General notation.}

For correctness, we use super- and subscript with values, sets and lists. A subscript \(i\) (e.g., \(r_i\)) refers to some element with index \(i\) in a set or list. A superscript \(i\) as in \(r^i\)  refers to the value \(r\) from party \party i. Likewise \(R^i\) refers to the list or set from party \party i and \(r^i_j\) is the value with index \(j\) in the set from party \(i\).

When needed for clarity, values sent from party \party i to \party j, are denoted as \(r^{i\rightarrow j}\). 
% Furthermore, output sets from the ideal functionality are marked with a bar \(\bar{R}\). \jl{is this required?}

A collection of elements is marked as a capital \(E\).
$|E|$ denotes the cardinality of the collection (i.e.,  its size). %To avoid confusion, the bit length of a value $v$ is denoted by $\len(v)$.
A set is an unordered collection with elements of the same type denoted using curly brackets: $\{e_1, \ldots, e_{|\set{}|}\}$. 
Unlike sets, vectors maintain the order of the elements and are enclosed by angle brackets: $\vect=\langle e_1, \ldots, e_{m} \rangle$, with $m$ the number of elements in the vector.
Tuples also maintain the order of elements, but have elements of different types and are enclosed by parentheses: $(e_1, \ldots, e_{m})$.

\noindent $[m..n]$ denotes the set of positive integers from $m$ up to $n$, and $[n]$ is a shorthand for the set of positive integers from $1$ up to $n$.
A uniform random selection over all strings of bit length $l$ is denoted by $r \xleftarrow[]{\$} \{0,1\}^l$ and the \emph{exclusive or} (xor) operation by $\oplus$.

\paragraph{Symmetric Encryption.}

A \emph{ symmetric encryption scheme} secure against attacks targeted against the indistinguishability of ciphertexts (IND-CPA), consists -- slightly simplified -- of two algorithms: \enc and \dec.
Algorithm \enc is defined by $\enc_n: \{0,1\}^n \times \{0,1\}^{l_m} \rightarrow \{0,1\}^{l_c}$ and takes as input a key of length $n$ and a message of arbitrary size $l_m$. The output (of size $l_c$) is called the ciphertext. Algorithm \dec is the corresponding decryption algorithm defined by $\dec_n: \{0,1\}^{n}\times \{0,1\}^{l_c} \rightarrow \{0,1\}^{l_m}$. It takes the same key as input together with the ciphertext and turns it into the corresponding plaintext.

\paragraph{Pseudorandom Permutation (PRP).}
A (keyed) \emph{pseudorandom permutation}
is defined by $\prp{n}: \{0,1\}^n \times \{0,1\}^n \rightarrow \{0,1\}^n$, which takes as input a key of length $n$ and an input block of size $n$. %The output of size $n$ is called the ciphertext.
A pseudorandom permutation is secure in the standard model if a distinguisher $\distinguisher$ cannot distinguish $\prp{n}$ from $\Perm_n$, which is the set of all permutations on $\{0,1\}^n$. More formally, for any probabilistic polynomial time (PPT) algorithm $\distinguisher$, there is a negligible function $\negl$ such that:
$|Pr[\distinguisher(\prp{n}(k, .),\iprp{n}(k,. )) = 1]
- Pr[\distinguisher(f(.),f^{-1}(.)) = 1]| \le \negl(n),$
where $k$ and $f$ are uniform random choices over $\{0,1\}^n$ and $\Perm_n$ respectively \cite{K14}.

\paragraph{Threshold Secret Sharing.}
\emph{$(t,n)$ Threshold Secret sharing} \cite{B11} was introduced by Shamir \cite{S79} and Blakley \cite{B79} and allows to split a secret $k$ into $n$ pieces, called shares, such that $t$ (the threshold) of these shares are required to correctly reconstruct the initial secret. With less shares, no information about the secret can be learned.

Let $\tsssplit{t,n, \kappa}: \{0,1\}^{\kappa} \rightarrow \mathbb{S}^{n}$ and $\tssrecon{t, \kappa}: \mathbb{S}^{t} \rightarrow \{0,1\}^\kappa$ denote a pair of functions; the first splits a secret of length $\kappa$ into $n$ shares with threshold $t$, while the second reconstructs the original secret based on $t$ shares.
The domain of secret shares for the function pair $(\tsssplit{t,n,\kappa}, \tssrecon{t,\kappa})$ is denoted by $\mathbb{S}$.

A threshold secret sharing scheme is perfectly secure if a distinguisher, which is given less than $t$ shares, is unable to derive information about the secret.
Formally,
$\forall~t,n \in \mathbb{N}_0$ with $t\le n$, $\forall~ k_0, k_1 \xleftarrow[]{\$} \{0,1\}^{\kappa}$, ${S}_i \leftarrow \tsssplit{t,n,\kappa}(k_i)$ for $i \xleftarrow[]{\$} \{0,1\}$, $\forall ~{S_j} \subseteq {S}_i$ with $|{S_j}| < t$: $Pr[\distinguisher({S_j},k_0) = 1] = Pr[\distinguisher({S_j},k_1) = 1]$.

In Shamir's proposal, each element of $\mathbb{S}$ is a tuple of the form $(x,y)$, which represents a point on a polynomial of degree $t-1$ over a finite field. All $x$ values can be taken from $[n]$ to reduce their bit length to $\lceil \log_2(n) \rceil$.

% TODO: Shamir’s scheme is considered information-theoretically secure: REF
% https://ethresear.ch/t/security-considerations-for-shamirs-secret-sharing/4294

%https://crypto.stackexchange.com/questions/39970/shamirs-secret-sharing-scheme-prime-security
\paragraph{Oblivious Key Value Store (OKVS).}

An \emph{Oblivious Key Value Store} is a data structure, formalized by Garimella et al.~\cite{GPRTY21}, that compactly represents a key-value mapping $k_i \mapsto v_i$. When the values are randomly chosen over a uniform distribution, the OKVS hides the keys $k_i$ that were used to generate it.
%More formally, a key-value store, parameterized by a set $\mathbf{K}$ of keys and $\mathbf{V}$ of values, consists of the algorithms $\encode$ and $\decode$. $\encode$ takes a list of key-value pairs $(k,v)\in L$ and outputs an object $\mathbf{S}$, (or $\bot$ with statistically small probability). $\decode$ takes as input an object $\mathbf{S}$, and a key $k$, and outputs a value $v$.
%Reference: https://dl.acm.org/doi/pdf/10.1145/3460120.3484772?casa_token=T5GZtzWePzgAAAAA:f2ZYwH-1jFq6DeQNIVNmKP0D_YpxDP1BeEZa9P-vnsvtgokC732oAGLWUwQEpq06IbBeSzSaPwfCMxg
\begin{definition}
  A {\em\bf Key-Value Store (KVS)} is parameterized by a set of keys $\mathcal{K}$ and a set of values $\mathcal{V}$, and consists of two algorithms:
  \begin{itemize}
    \item $\encode{}$ takes as input a set of key-value pairs ($k_i, v_i$) and outputs, with non-negligible probability, a data structure $\mathcal{S}$ or an error indicator $\bot$ otherwise. Optionally, parameters can be passed as additional input.
    \item $\decode{}$ takes as input the data structure $\mathcal{S}$ and a key $k$, and outputs a value $v$.
  \end{itemize}
  A {\em KVS} is {\bf correct} if, for all $A \subseteq \mathcal{K}\times\mathcal{V}$ with distinct keys:
  $(k,v) \in A$ and $\bot \ne \mathcal{S} \leftarrow \encode{}(A) \Rightarrow \decode{}(\mathcal{S},k) = v$.
\end{definition}
\begin{definition}
  \noindent A {\em KVS} is an {\em\bf Oblivious Key-Value Store (OKVS)}, if for all distinct keys $\{k^0_1,...,k^0_n\}$ and all distinct $\{k^1_1,...,k^1_n\}$, if $\encode{}$ does not output $\bot$ for $\{k^0_1,...,k^0_n\}$ and $\{k^1_1,...,k^1_n\}$, then the output of $\mathcal{R}(k^0_1,...,k^0_n)$ is computationally indistinguishable from that of $\mathcal{R}(k^1_1,...,k^1_n)$, where $\mathcal{R}(k_1, ..., k_n)$ is defined as:
  \begin{itemize}
    \item For $i \in [n]$, $v_i \xleftarrow{\$} \mathcal{V}$.
    \item Output $\encode{}$($\{(k_i, v_i)\}_{i\in[n]}$).
  \end{itemize}
\end{definition}

\noindent In other words, if the OKVS encodes random values, it is infeasible to tell whether any key $k \in \mathcal{K}$ was used to generate $\mathcal{S}$ or not.\\ In the remainder, we annotate output of an OKVS with a tilde (e.g., $\widetilde{v_i} = \decode{}(\mathcal{S},k_i)$), denoting that the receiver is oblivious of whether or not the key-pair $(k_i,v_i)$ was encoded in $\mathcal{S}$.

Several OKVS constructions exist of which a polynomial $p$, that is chosen using interpolation such that $p(k_i) = v_i$, is the simplest and size-optimal. Others use Garbled Bloom Filters~\cite{CLZ13}. Pinkas et al. define an efficient OKVS, called a {\it probe and XOR of strings (PAXOS)}~\cite{PRTY21}. Garimella et al. made further improvements to the PAXOS scheme defining 3H-GCT and Bienstock et al. present a near optinal OKVS based on random band matrices~\cite{BPSY23}.

In our implementation, we use the construction of Raghuraman and Rindal~\cite{raghuraman2022blazing}.

\begin{definition}[Random decodings]
  An OKVS satisfies \emph{random decodings} if, for all sets of \( n \) distinct keys \( A = \{ k_1, \dots, k_n \}
  \subset \mathcal{K} \), \( n \) values \( \{ v_1, \dots, v_n \} \) each drawn uniformly at random from \(
  \mathcal{V} \), the output of \( \decode{}(\mathcal{S}, k) \) for a key \( k \not\in A \) is statistically
  indistinguishable from a uniformly random element in \( \mathcal{V} \), where \( \mathcal{S} \gets \encode{} \left(
  \{ (k_1, v_1), \dots, (k_n, v_n) \} \right) \).
\end{definition}

\paragraph{Security Model.}
In our multi-party setting, we have two security models. The colluding model for the data providers, where we allow any $n-1$ of the $n$ data providers to collude, and the non-colluding model for the data collector, who is not allowed to collude with any data provider.
The adversary model used in this work is the semi-honest model. In the semi-honest model, it is assumed that the parties follow the protocol as prescribed, but they may learn and retain information about the inputs and intermediate values.

%\section{Practical M-PSI}\label{sct:mpsi}

\section{The Set Intersection Key Agreement Functionality and Protocol}
\label{sct:sika}
To support the requirements outlined in Section~\ref{sec:req} and construct our Labeled D-PSI protocol (Section~\ref{sct:output}), we first present the Set Intersection Key Agreement protocol (SIKA), a composable building block suitable for implementing several practical applications.

The ideal functionality of this protocol is formalized in Functionality~\ref{func:core}~\Fcore. There are $n$ data providers (data sources) \( \party 1,\ldots, \party n \) with input sets \( \ID 1 ,\ldots, \ID n \) of equal size $m=|{\ID
i}|$ and a single data collector \thirdparty. From the functionality each data provider learns, for each entry in its local
input set, a secret key and a corresponding pseudonym. For all the entries in the intersection of the input sets, and
only those entries, \thirdparty learns from the functionality the same secrets keys and pseudonyms and is able to link
those across the different input sets through the use of $k$.
%The release of the secrets and pseudonyms is to allow the core functionality to be used as a building block in combination with different output protocols, presented in Subsection~\ref{sct:output}.

Protocol~\ref{prot:core} ~\Pcore realizes this ideal functionality. In the protocol, the $n$ data providers
each generate shares for each identifier \id i k in their local input set \ID i (step 1). Those shares are used to
jointly compute blinded pseudonyms provided to each \party i (step 2 and 3). If an \id i k is in the
intersection, each \party i obtains a (data provider specific) blinding of the same jointly computed pseudonym. However, if
it is not in the intersection, each \party i obtains a blinding of a different random value. To realize this, and ensure
that data providers do not learn information about the input set of other parties, the protocol makes use of {\em Oblivious Key
Value Stores} to share the blinded shares. Hence, if the identifier is not in the intersection, at least one data provider \party
i has no corresponding blinded share included in his \okvs i j for each of the other receiving parties \party j. As a result, each
\party j computes a different pseudo-random value, resulting in a blinded random, and unlinkable value \bosecret j k.
Finally, the data providers send their blinded pseudonyms to the data collector \thirdparty, together with extra information to allow for unblinding the pseudonym (the $z$-values). If and only if the corresponding \id i k
is in the intersection, \thirdparty is able to link the unblinded jointly computed global pseudonym
\secret k (step 4). For each \id i k, the protocol also computes a secret key $sk^i_k$ which can be used by $P^i$ to encrypt a possible payload. Every data source does this based on the $z$-values. The collector is only able to reconstruct this key for nyms in the intersection. For others, it has no clue what $z$-values to combine.

\begin{functionality}{\Fcore{} - Set Intersection Key Agreement}{func:core}

  {\sc Parameters:} security parameter $\kappa$, the size of the inputs $m$, $n$ data providers \party 1, \ldots, \party n, and data collector \thirdparty.

  {\sc Input:} each \party i has input ${X}^i = \{id^i_1,\ldots, id^i_m \}$\footnote{For ease of presentation, all ${X}^i$ have cardinality $m$.}, with $id^i_k \in \{0,1\}^\kappa$,\\
  \thirdparty has no input.

  {\sc Functionality:}
  \begin{enumerate}
    \item[(0)] Let \parties = $\emptyset$
    \item Upon receipt of $({\tt input}, {X}^i)$ from each data provider:\\
      \-\quad- Let $ X_\cap = \bigcap_{j\in [n]}{X}^j$\\
      % Original specification:
      %\-\quad- Let ${Y}_\cap = \left\{(id,k)\mid id\in  X_\cap,\ k \in \mathbb{N}_{0},\ 1 \leq k \leq | 
      %X_\cap|,\ k \mbox{ unique} \right\}$.\\
      % Better?
      \-\quad- Let ${Y}_\cap = \left\{(id,k)\mid id\in  X_\cap,\ k = index({X}_\cap, id) \right\}$.\\
      \-\quad- For each data provider \party i:\\
      \-\qquad \comments{  // associate random nym and key with each identifier}\\
      \-\qquad- Let ${M}^i = \left\{(id, \bsecret{i}{}, sk^i)\mid id\in {X}^i,\ (\bsecret{i}{}, sk^i)\xleftarrow{\$}\{0,1\}^{\kappa\times2}\right\}$.\\
      \-\qquad- Send ${M}^i$ to \party i.
    \item Upon receiving $({\tt proceed})$ from a data provider \party i and ${Y}_\cap\neq \emptyset$:\\
      \-\quad - Add \party i to  \parties.\\
      %\-\quad- Let $\bar{\result}^i = \emptyset$.
      \- \quad- If $|\parties|=n$, with all $(id,\bsecret{i}{} , sk^i)\in {M}^i$, create $\result^i$ as follows, and add it to $\result$:
      \begin{align*}
        \result^i=\left\{
        \begin{cases}
          (\bsecret {i} {},k, sk^i)       & (id, k)\in {Y}_\cap        \\
          (\bsecret {i} {},\ \bot,\ \bot) & id \notin {X}_\cap \\
        \end{cases}\right\}
      \end{align*}

      % \-\quad- For each $\bsecret {} k \in B$:\\
      % \-\qquad If $(id,\bsecret {} k,sk) \in \bar{M}^i$ and $(id,\secret k)\in \bar{Y}_\cap$:\\
      % \-\quad\qquad Add $(\secret k, sk)$ to $\bar\result^i$.\\
      % \-\qquad Else add $(r,sk)$ to $\bar\result^i$, with $r\xleftarrow{\$}\{0,1\}^\kappa$.\\
      \-\qquad- Send $\result$ to \thirdparty.
      % \-\quad- Let $\bar{N^i} = \left\{\bsecret i k\mid (.,\bsecret i k, .)\in \bar{M}^i\right\}$.\\
      % \-\quad- Let $\bar{\result} = \left\{ (bnym^1, sk^1_{}, \ldots, bnym^n, sk^n_{}) \mid id_\cap \in X_\cap,\forall i\in[n] : (id_\cap,bnym^i_{},sk^i_{})\in \bar{M}^i \right\}$.\\
      % \-\quad- Send $\bar{N}^1,\ldots,\bar{N}^n$ and $\bar{\result}$ to \thirdparty.
  \end{enumerate}
  \label{func:core2}
\end{functionality}

\begin{protocol}{\Pcore}{prot:core}%
  {\sc Parameters:} security parameter $\kappa$, the size of the inputs $m$, $n$ data providers \party 1, \ldots, \party n, and data collector \thirdparty.

  {\sc Input:} each \party i has a set $X^i$ of identifiers $id^i_k \in \{0,1\}^\kappa$ with input length $m$ \footnote{For ease of presentation, all $\ID i$ have cardinality $m$.}, \thirdparty has no input.

%  {\sc Output:} each \party i gets a pseudonym $\bosecret i k$ and secret $sk^i_k$ for each $id^i_k\in X^i$, \thirdparty computes for each $id$ the corresponding pseudonym, and for entries $id_\cap$ in the intersection, the pseudonyms and secret $sk^i$ corresponding to the same $id_\cap$ of each party are linked.

  \noindent{\sc Protocol:}
  \begin{enumerate}
    \item Each \party i chooses a key $\key i\xleftarrow{\$}\{0,1\}^\kappa$, generates a random nym share $\share i k \xleftarrow{\$} \{0,1\}^\kappa$ for each $k \in [m]$, \\
      \- and random $\rnd i j k$-shares (of size $\kappa$) for each $k\in[m]$ and $j\in[n]$.
      \comments{\-\quad // {\em z}-values serve double purpose: } \\
      \-\quad \comments{// (a) they determine the secret key: $sk^i_k = \bigoplus_{j\in[n]} \rnd i j k$}\\
      \-\quad \comments{// (b) they are used in computing the blinding factors: $\prp{\kappa}(\key i, \rnd i j k)$}
    \item Each \party i sends to each other data provider \party j ($i\neq j$) an OKVS encoded as follows:\\
      \-\quad $\okvs i j =\encode m\biggl(\Bigl\{\bigl(id_k,\langle \bshare i j k, \rnd i j k\rangle\bigr) \Big| k\in[m], id_k \in \ID i,\bshare i j k \leftarrow  \share i k\oplus \prp{\kappa}(\key i, \rnd i j k)\Bigr\}\biggr)$.\\
      \-\quad \comments{// $\rnd i j k$ is encoded in the OKVS, because C needs it to remove part of the blinding and to recover $sk^i_k$}
    \item Every \party j , once it received an $\okvs i j$ from all other data providers:\\
      - Recover all $bs$ and $z$-values from $\langle \boshare i j k, \ornd i j k \rangle \leftarrow \decode m(\okvs i j, id^j_k)$ for each $id^j_k\in X_j$ and $\okvs i j$ received from each \party i.\\
      - Set $M^j = \bigl\{(id_k,\bosecret j k, sk^j_k)\big|k \in [m],id_k\in X_j,\bosecret j k = \share j k \bigoplus_{i\neq j} \boshare i j k, sk^j_k=\bigoplus_{i\in[n]}\rnd j i k\bigr\}$.\\
      - Send \key j and $B^j=\bigl\{(\bosecret j k,\langle\ornd 1 j k,\ldots, \ornd n j k\rangle ) \big| k \in [m]\bigr\}$ to \thirdparty.\\
      - Output $M^j$.

    \item Upon receipt of $B^i$ with entries $(\bosecret i k,Z^i_k)$ and \key i from data provider \party i, \thirdparty computes the following:\\
      - If \party i provided input before: Ignore $B^i$ and \key i.\\
      - If every \party j provided $B^j$:\\
      \-\quad- Let $\result=\emptyset$.\\
      \-\quad- For each \party i:\comments{ // map blinded pseudonym to global pseudonym by removing the blinding}\\
      \-\qquad- Let $N^i=\bigl\{(\osecret i k,\bosecret i k)\big|k\in[m]\bigr\}$ with $\osecret i k \leftarrow \bosecret i k  \bigoplus_{j\neq i} \prp{\kappa}(\key j, \ornd j i k)$.\\
      \-\quad \comments{//compute intersection: If the same global $nym$ is found in $N^1$ and $N^2$, it must be found in all $N^i$}\\
      \-\quad - Let $Nyms_\cap = \bigl[\osecret {} {}\big| (\osecret {} {},.,.) \in N^1, (\osecret {} {},.,.) \in N^2 \bigr].$ \\
      \-\quad- For each \party i:\\
      \-\quad\quad \comments{// Tricky! Recover the secret key $sk^i_k$, by retrieving all the $\ornd i j k$ from $B^j$, with $j \in [n]$.}
      \begin{align*}
        \-\quad\quad\quad \result^i=\left\{
        \begin{cases}
          (\bosecret {i} {k},p_k, sk^i_k)   & (\osecret i k, \bosecret i k) \in N^i \text{ \bf and } \osecret i k \in Nyms_\cap,  \text{ with } p_k = index(Nyms_\cap, \osecret i k) \text{ and  }                                                                   \\
          & sk^i_k = \bigoplus_{j\in[n]} \ornd i j k \text{ where } ((\osecret i k, \bosecret j k) \in N^j  \text{ \bf and } (\bosecret j k,\langle\ldots\ornd i j k\ldots\rangle \in B^j)) \\
            (\bosecret {i} {k},\ \bot,\ \bot) & (\osecret i k,\bosecret i k) \in N^i \text{ \bf and } \osecret i k\notin Nyms_\cap.                                                                        
          \end{cases}\right\}
        \end{align*}
        \-\quad\quad \text{\ Add } $R^i$ \text{ to } R.\\
        \-\quad - Output $\result$.

  \end{enumerate}
\end{protocol}

\paragraph{\bf Correctness.}
We first show that, in the absence of corruptions, the outputs of both ideal and real protocol are indistinguishable.
%In the ideal functionality, each input party \party i outputs a set of records containing for each $id$ in its input set the randomly generated nym and secret. The third party \thirdparty obtains for each and only for each $id$ in the intersection $X_\cap$ of all input sets, a single pseudonym and from each input party the corresponding nyms and secrets.
As in the ideal functionality, the real data provider \party i obtains a set $M^i$ with random $\bsecret {} {}$ and random
$sk$ for each of its \(k\) \(id\)'s. Both are indistinguishable from random as they are, respectively, the XOR of local
randomness and the XOR of the (partial) outputs of the OKVSs from all other data providers. For $bnym$ this partial
output is the first $\kappa$ bits, while $sk$ is based on the last $\kappa$ bits in the OKVSs.

To show that the output of the data collector is also correct, only if $id$ is in the intersection, we see that \thirdparty
is able to link the blinded $bnym$ of each data provider to the same nym. We distinguish two cases:
\begin{enumerate}
  \item The $id$ is in the intersection and each data provider obtains a value which was inserted into the OKVS from the
    other parties. In this case, the blinded $bnym$ is the XOR of a jointly computed $nym$ (which is the same for each
    \party i) blinded with pseudorandom permutations of known $z$-values. The data collector is able to unblind the $nym$
    using the keys and $z$-values received from the data providers, link the records and reconstruct the secret key \( sk
    \).
  \item The $id$ is not in the intersection; for this $id$, each $bnym$ is the result of an XOR with at least one value
    indistinguishable from random (by definition of an OKVS for values not encoded), hence, unlinkable.
\end{enumerate}

% From the preceding high-level description of \Pcore, it is clear that the protocol is correct with respect to the ideal functionality \Fcore except in the event of a false positive — i.e., $\secret k = \bsecret 1 k  \bigoplus_{j\neq 1} \prp{\kappa}(\key 1, \rnd 1 j k) = \ldots = \bsecret n k  \bigoplus_{j\neq n} \prp{\kappa}(\key i, \rnd n j k)$ for some $\id i k \in \ID i$ not in the intersection. Let \party i be the party that did not have \id 1 k in its input set. That party will not have programmed its OKVSs with that \id 1 k.
% As a result, \rnd 1 i k, \boshare 1 i k, and consequently, \bosecret 1 k will be pseudorandom.
% The probability that the unblinded \osecret 1 k matches the other $n-1$ unblinded \osecret j k is negligible.

From the preceding high-level description of \Pcore, it is clear that the protocol is correct with respect to the ideal
functionality \Fcore except in the event of a false positive, i.e., $\secret k = \bsecret 1 k \bigoplus_{j\neq 1}
\prp{\kappa}(\key 1, \rnd 1 j k) = \ldots = \bsecret n k  \bigoplus_{j\neq n} \prp{\kappa}(\key n, \rnd n j k)$ for some
$\id {} k \in \ID i$ and $\id {} k \in \ID j$ for $j\in[n]\setminus\{i\}$. Since \party i does not have \id {} k in its
input set, it will not have programmed its OKVS with this \id {} k. As a result, \ornd i j k, \boshare i j k, and
consequently, \bosecret j k (for each other \party j) will be pseudorandom. The probability that the unblinded \osecret
i k matches the other $n-1$ unblinded \osecret j k is negligible.

\paragraph{\bf Security.}

\begin{theorem}
  \label{thm:core_protocol_security}
  Protocol~\Pcore is secure in the semi-honest model, against any number of colluding, semi-honest input
  parties and against a non-colluding, but possibly malicious data collector.
\end{theorem}

% To prove formally the security in the standard model in case of semi-honest participants, we prove that the view of
% each party in the protocol execution can be simulated, given its input and output. This implies that the party learns
% nothing from the protocol execution beyond what it can derive from its inputs and prescribed output~\cite{L17}.
% Therefore, we create simulators and prove that the view resulting from interaction with the real protocol, denoted by
% \Pcore, and the view resulting from interaction with the simulator can be distinguished only with a negligible
% probability. The view consists of the input, the party’s internal random tape and the received messages. The view
% created by the simulator must be consistent with its given output.

\begin{proof}
  We first prove that the protocol is secure in the presence of semi-honest colluding data providers, followed by a proof
  in the presence of a semi-honest data collector.

  \paragraph{Consider $n-1$ Semi-honest Data Providers.}
  Colluding entities share everything with each other: inputs, outputs, messages received and sent, and random values
  used. We now prove that the protocol is secure against $n-1$ colluding semi-honest data providers. As a consequence, it
  is also secure against smaller sets of colluding semi-honest data providers. Since $n$ data providers already have all
  the information, allowing $n$ colluding data providers is senseless. Without loss of generality, let \party 1 be the
  only uncorrupted data provider. Let \parties denote all the data providers and \Pcor all corrupted data providers (i.e.,
  $\parties\setminus \{\party 1\}$).

  \begin{simulation}{$\simulator_{\Pcor}$}{sim_core}
  \item For corrupted \( \party i \in \Pcor \), \simulator receives \( X^i \) from \environment. It then sends \( ({\tt
    input}, X^i) \) to \Fcore{} and receives \( M^i = \left\{ (id^i_{k}, bnym^i_{k}, sk^i_{k}) \mid id^i_{k} \in X^i
    \right\}. \)
    % Sample the PRP key according to the protocol.
    For every \( \party{i} \in \parties \), \simulator{} samples \( \key{i} \xleftarrow{\$} \{0,1\}^\kappa \) and initializes an empty list~\( \List_{id} \).

  \item For \( i \in [2, n] \):
    \begin{enumerate}
        % Compute the z-shares of the corrupted parties based on the sk from the functionality.
      \item For each \( j \in [n] \setminus {i} \) and for \( k \in [m] \), \( \simulator \) uniformly samples \(
        z_k^{i \to j} \) and then computes \( z^{i \to i}_k \gets sk^i_k \oplus \bigoplus_{j \in [n] \setminus \{ i
        \}} z^{i \to j}_k \).

        % Compute the OKVSs sent by the corrupted parties according to the protocol.
      \item For \( j \in [n] \setminus {i} \), \( \simulator \) computes \( okvs^{i \to j} \) according to the protocol.

      \item For \( k \in [m] \), \( \simulator \) checks whether there exists some \( s^\ast \) such that \( (id_k,
        s^\ast) \in \List_{id} \) for \( id_k \in X^i \). If it does, \( \simulator \) sets \( s^{1 \to i}_k \gets
        s^\ast \).
        Otherwise, \( \simulator \) samples \( s^{1 \to i}_{k} \) uniformly at random and adds \( (id_k, s^{1 \to
        i}_{k}) \) to \( \List_{id} \).

        % Compute the z-shares of the honest party so that the corrupted parties produce the correct bnyms
      \item For \( k \in [m] \), \( \simulator \) retrieves \( (id_k, s^{1 \to i}_{k}) \), for \( id_{k} \in X^i \) from
        \( \List_{id} \) and computes
        \[
          z^{1 \to i}_k \gets PRP^{-1} \left( key_1,
            \begin{array}{c}
              bnym^i_{k} \oplus s^i_{k} \oplus s^{1 \to i}_{k} \\
              \oplus \bigoplus_{j \in [n] \setminus \{ 1, i \}} bs^{j \to i}_{k}
          \end{array} \right),
        \]
        where \( bnym^i_{k} \) comes from \( M^i \) and \( bs^{j \to i}_{k} \) is decoded from \( okvs^{j \to i} \)
        according to the protocol.

      \item \( \simulator \) then constructs \( okvs^{1 \to i} \) using \( id_{k} \) from \( X^i \) and the
        corresponding \( s^{1 \to i}_k \) and \( z^{1 \to i}_k \) as generated above.
    \end{enumerate}

  \end{simulation}

  Simulator~\ref{sim_core}, denoted \( \simulator_{\Pcor} \), runs the real adversary \adversary internally and passes
  the messages between the environment \environment and \adversary. Likewise, it passes the messages (e.g., leakage of
  state and continuation) between the semi-honest parties and~\adversary.

  To demonstrate the correctness of this simulation, we describe the following sequence of hybrid simulators which
  behave incrementally different from the real protocol until the final hybrid is the simulator \( \simulator_{\Pcor}
  \). Showing that each change is indistinguishable from the previous one, from the perspective of the environment \(
  \environment \), demonstrates that the real execution of the protocol is indistinguishable from the ideal execution of
  the \( \Fcore \) functionality.

  \textbf{Hybrid 1.}
  For each simulated \( \party{i} \), and for \( k \in [m] \), \( \simulator_\Pcor \) first samples \( sk^i_k \) at random, then
  samples \( z^{i \to j}_k \) at random for \( j \in [n] \setminus \{ i \} \), and then computes
  \[
    z^{i \to i}_k \gets sk^i_k \oplus \bigoplus_{j \in [n] \setminus \{ i \}} z^{i \to j}_k.
  \]
  This is identically distributed to the real protocol.

  \textbf{Hybrid 2.}
  The simulator initializes an empty list \( \mathcal{L}_{id} \). For \( i \in [1, n] \) and for \( k \in [m] \) the
  simulator first checks if \( id^i_{k} \in X_i \) is saved in \( \mathcal{L}_{id} \); if it is, the simulator sets \(
  s^{1 \to i}_{k} \) according to the list. If \( id^i_{k} \) is not in the list, the simulator samples \( s^{1 \to
  i}_{k} \) uniformly at random, and then adds \( (id^i_{k}, s^{1 \to i}_{k}) \) to the list \( \mathcal{L}_{id} \).
  This method ensures that if any identifier is present in multiple input sets \( X_i \), then the same \( s^{1 \to i}
  \) value is assigned to it.

  The simulator then constructs the \( n - 1 \) OKVSs \( okvs^{1 \to i} \) using the identifiers \( id^1_{k} \in X_1 \),
  for \( k \in [m] \), as in the real protocol, but with the \( s^{1 \to i}_{k} \) values as determined by \(
  \mathcal{L}_{id} \) for these identifiers. The consequence of this is that the OKVSs sent by party \party 1 to parties \party i
  and \party i' will be based on different \( s^{1 \to i}_k \) and \( s^{1 \to i'}_k \) values for the same identifier \(
  \id 1 k \) if that identifier is not in \( X_i \cap X_{i'} \).

  Let \( G \) denote the game played by the environment against Hybrid~1, and let \( G' \) denote the game it plays
  against Hybrid~2. Furthermore, let \( \Delta(G, G') \) denote the difference in the probability that \( \environment
  \) outputs 1 when playing either \( G \) or \( G' \). First, we observe that \( G = G \mid PRP \), and the same holds
  for \( G' \) since the real protocol uses a \( PRP \) for the inputs to the OKVSs. We then have the following:
  {\small
  \begin{align*}
    \Delta(G \mid PRP, G' \mid PRP) \leq {}              & 2 \Delta(PRP, PRF)                                   \\
    & {} + \Delta(G \mid PRF, G' \mid PRF)                 \\
    \Delta(G \mid PRF, G' \mid PRF) \leq {}              & 2 \Delta(PRF, \mathcal{U})                           \\
    & {} + \Delta(G \mid \mathcal{U}, G' \mid \mathcal{U}) \\
    \Delta(G \mid \mathcal{U}, G' \mid \mathcal{U}) = {} & 0,
  \end{align*}}%
  where \( \Delta(PRP, PRF) \) and \( \Delta(PRF, \mathcal{U}) \) respectively denote the distinguishing error between a
  PRP and a PRF, and between a PRF and the uniform distribution (assuming the multi-query versions of the PRP-PRF
  distinguishing experiment and of the PRF security experiment).

  \textbf{Hybrid 3.}
  For \( i \in [2,n] \), each \( P_i \) first samples \( bnym^i_{k} \) at random, for \( k \in [m] \), and \( P_1 \)
  then computes {\small
    \begin{align*}
      z^{1 \to i}_{k} & \gets PRP^{-1} \left( key_1,
        bnym^i_{k} \oplus s^i_{k} \oplus s^{1 \to i}_{k} \oplus
      \bigoplus_{j \in [n] \setminus \{ 1, i \}} \widetilde{bs}^{j \to i}_{k} \right). \\
    \end{align*}
  } Since \( bnym^i_{k} \) is uniformly distributed, so will be the input to \( PRP^{-1} \) and so will be \( z^{1 \to
  i}_{k} \), identically to how it was sampled in the protocol. The last value \( z^{1 \to 1}_{k} \) does not need
  to be computed since it remains outside of the environment's view because \( \party{1} \) and the data collector \thirdparty are both honest. Only if an identifier is in the intersection, then the environment can reconstruct \( z^{1 \to
  1}_{k} \) from the other \( z^{1 \to i}_{k} \) shares and the \( sk^1_k \) value returned by the data collector, but
  this would still be identically distributed to the real protocol.

  \textbf{Hybrid 4.}
  The simulator now computes the list \( \mathcal{L}_{id} \) using only \( i \in [2, n] \), i.e., it does not populate
  it with information from \( X^1 \). To build the OKVSs, the simulator now uses the identifiers \( id^i_{k} \in X_i \)
  for \( okvs^{1 \to i} \), together with the \( s^{1 \to i}_{k} \) values from the list \( \mathcal{L}_{id} \) and the
  \( z \)-shares \( z^{1 \to i}_{k} \).

  For each \( okvs^{1 \to i} \), the differences with Hybrid~3 are:
  \begin{enumerate}
    \item the entries with identifiers \( id^1 \in X^1 \setminus (X^1 \cap X^i) \) are no longer encoded, and
    \item the entries with keys \( id^i \in X^i \setminus (X^1 \cap X^i) \) are now encoded.
  \end{enumerate}
  Denoting by \( G \) the game against Hybrid~3, and by \( G' \) the game against Hybrid~4, we then have the following:
  \begin{align*}
    \Delta(G \mid PRP, G' \mid PRP) \leq {}              & 2 \Delta(PRP, PRF) + 2 \Delta(PRF, \mathcal{U})       \\
    & {} + \Delta(G \mid \mathcal{U}, G' \mid \mathcal{U}), \\
    \Delta(G \mid \mathcal{U}, G' \mid \mathcal{U}) = {} & \epsilon_{oblv}.
  \end{align*}
  where \( \epsilon_{oblv} \) denotes the obliviousness error of the OKVS.

  \textbf{Hybrid 5.}
  The simulator no longer samples but instead obtains the \( sk^i_{k} \) and \( bnym^i_{k} \) values for the semi-honest
  parties within the \( M^i \) sets from the functionality; the view that this presents to \( \environment \) is
  identically distributed to that of Hybrid~4.

  \paragraph{Consider a malicious but non-colluding data collector.}

  The simulator \( \TPSimulator \) receives \( \TPInputSet = (\TPInputSet^1, \dots,
  \TPInputSet^n) \) from \( \Fcore \), with \( |\TPInputSet^i| = m \) and
  \[
    \TPInputSet^i =\left\{
    \begin{cases}
      (bnym^i_j, k, sk_j^i), & (id_j, k) \in \IntersectionInfoSet, \\
      (bnym^i_j, \bot, \bot), & (id_j, \ast) \notin \IntersectionInfoSet.
    \end{cases}\right\}
  \]
  The malicious but non-colluding data collector outputs to the environment \( \environment \) the same set \( \TPInputSet \) that it
  received from \( \Fcore \). Therefore, \( \TPSimulator \)'s only task is to simulate the messages sent from
  each honest data provider \( \party{i} \) to \( \thirdparty \) identically to the protocol. (This is assuming that the honest
    parties use private and authenticated channels to communicate with each other so that the adversary cannot observe the
  messages sent between honest parties.)

  To simulate \( key_i \), the simulator samples it uniformly at random as in the protocol. To simulate \(
  \bosecret{i}{j} \), for \( j \in [m] \), \( \TPSimulator \) uses the \( bnym^i_j \) from \(
  \TPInputSet^i \). To simulate the \( z \)-shares sent by each \( \party{i} \), the simulator samples \( z^{2
  \to i}_j, \dots, z^{n \to i}_j \) uniformly at random and sets
  \[
    z^{1 \to i}_j =
    \begin{cases}
      sk^i_j \oplus \bigoplus_{l = 2}^{n} z^{l \to i}_n & (\ast, \ast, sk^{i}_{j}) \in \TPInputSet^i, \\
      \stackrel{\$}{\gets} \{0, 1\}^\kappa & \mbox{otherwise}.
    \end{cases}
  \]
  This sampling of the \( z \)-shares is indistinguishable from the real protocol's distribution since their correlation
  with the \( sk \) key is ensured when the corresponding identifier is in the intersection of the input sets and since
  the environment has no information about the honest parties' real values of the \( z \)-shares under the assumption of
  private channels.
\end{proof}

\section{Output Protocols}
\label{sct:output}
Based on the SIKA protocol, we present a number of practical output protocols, including the Labeled D-PSI with payload, that show the flexibility of the construction.
We build upon universal composability to derive the security of the combined core and output protocols.
Since most output protocols are trivial w.r.t. the SIKA protocol, the proofs of the combined protocols are omitted; we include an argument for the security of the main payload protocol based on~\Fcore.

\subsubsection*{sub1 - Cardinality} A simple result is to obtain the size of the intersection: the data collector outputs the highest value of $k$ in its result set \result.

\subsubsection*{sub2 - Labeled D-PSI}
Many practical scenarios (including that of Section~\ref{sct:lethelink}), require the inclusion of a payload.
Applications often require revealing records associated with identifiers in the intersection. For privacy reasons, the
identifiers, even the ones in the intersection, should remain hidden.

Therefore, we develop an output protocol that supports payloads. While standard (D-)PSI requires to first obtain the intersection, followed by a privacy-friendly protocol to retrieve the records, our approach is different. The payload is encrypted in advance for each entry in the local input set, and passed to the data collector,  without knowledge of the intersection. The encryption ensures that the data collector can only decrypt a record if its identifier is in the intersection, thereby protecting the confidentiality of non-intersecting entries.
To implement this $(n,n)$-threshold encryption of the payload, we use the keys given to the data providers for each entry in their input set and the keys received by the data collector for each record in the intersection.
In order to encrypt a record with identifier \id i k, each \party i takes its encryption key $\tilde sk^i_k$ and encrypts the payload corresponding to this identifier. The ciphertexts are then sent to \thirdparty together with the corresponding local pseudonym (i.e., \bosecret i k).
Each data provider \party i sends the encrypted records to the data collector \thirdparty:
\begin{align*}
  \Bigl\{\bigl(\bosecret i k, \enc_\kappa(\tilde sk^i_k,data^i_k)\bigr) \Big| k\in[m]\Bigr\}
\end{align*}

\begin{functionalityOne}{\Fpay}{func:pay}%
  {\sc Parameters:}\\
  \-\quad $\kappa$,\\
  \-\quad $n$ data providers \party 1, \ldots, \party n, and data collector \thirdparty.

  {\sc Input:} \\
  \-\quad Each \party i has input $\records i = \bigl\{(\id i k, \Atts i k)\bigr\}_{k\in [m]}$.\\
  \-\quad \thirdparty has no input.

  %{\sc Output:} \thirdparty outputs $\records \cap = \bigl\{(\Atts 1 k, \ldots,\Atts n k)\big| \id {} k \in \ID \cap\bigr\}$.

  {\sc Functionality:}
  \begin{enumerate}
    \item Upon receipt of \records i from each data provider \party i:\\
      \-\quad- Let $\ID i = \bigr\{\id i k|(\id i k, .)\in \records i\bigr\}$, and \\
      \-\quad- Set $\ID \cap = \bigcap_i \ID i$.\\
      \- Send $\records \cap = \bigl\{(\Atts 1 {k_1}, \ldots,\Atts n {k_n})\big|$\\
      \-\quad $ \id {} {} \in \ID \cap, \forall j\in[n]: (\id {} {},\Atts j {k_j})\in\records j\bigr\}$ to \thirdparty.
  \end{enumerate}

\end{functionalityOne}

\begin{protocolOne}{\Ppay}{prot:pay}%
  {\sc Parameters:}\\
  \-\quad \Fcore, $\kappa$,\\
  \-\quad $n$ data providers \party 1, \ldots, \party n, and data collector \thirdparty.

  {\sc Input:} \\
  \-\quad Each \party i has input $\records i = \bigl\{(\id i k, \Atts i k)\bigr\}_{k\in [m]}$.\\
  \-\quad \thirdparty has no input.

  \noindent{\sc Protocol:}
  \begin{enumerate}
    \item  Each data provider \party i (with input $\ID i= \bigr\{\id i k|(\id i k, .)\in \records i\bigr\}$), and the data collector \thirdparty (without input) engage in \Fcore.\\
      Every party \party i gets $\bar{M}^i$, and \thirdparty gets $\bar\result$.

    \item Every data provider \party i:\\
      \-\quad - For each (\id i k, \bsecret{i}{k} , $sk^i_k$) in $\bar{M}^ i$: \\
      \-\qquad- Let $c^i_k = \enc_\kappa(\tilde sk^i_k,\Atts i k)$.\\
      \-\quad- Send $\records i'= \{(\bsecret{i}{k}, c^i_k)\}_{k\in[m]}$ to \thirdparty.
    \item For the data collector \thirdparty,
      upon receipt of $ \records i'$ from \party i:\\
      \-\quad - Let $\records i^*\leftarrow \emptyset$.\\
      \-\quad - For each record $(\bsecret{i}{}, c^i)$ in $\records i'$,\\
      \-\qquad - If $(\bsecret{i}{}, k , sk^i)$ in $\bar\result$ and $sk^i_k\neq \bot$ :\\
      \-\qquad\quad- Let $\Atts i {} = \dec_\kappa(sk^i,c^i)$, and\\
      \-\qquad\quad - Add \( (k, \Atts i {}) \) to $\records i^*$.\\
      \-\quad Output $\records \cap = \bigl\{(k, \Atts 1 {k_1}, \ldots,\Atts n {k_n})\big|$\\
        \-\qquad\quad $ \exists~k\in[| X_i |],$\\
      \-\qquad\quad $ \forall j\in[n]: (k,\Atts j {k_j})\in\records j^*\bigr\}.$

  \end{enumerate}%\label{prot:pay}
\end{protocolOne}

Functionality~\ref{func:pay} and Protocol~\ref{prot:pay} present the ideal functionality \Fpay and its protocol implementation \Ppay, respectively. The protocol takes advantage of \Fcore. After running the ideal functionality, the data collector obtains the pseudonyms and corresponding secrets for each identifier in the intersection. Upon receipt of the ciphertexts and corresponding pseudonym, the data collector is able to find the corresponding $sk^i_k$ and decrypts the ciphertexts.

We briefly argue the security of \Ppay; since the simulator emulates \Fcore, it receives the input set of each \party{i} and samples the secret keys used for encryption.
Using these, it can perfectly simulate the messages sent to the data collector \thirdparty for the plaintexts contained in $\records{\cap}$ that it received from \Fpay.
For the other messages sent from \party{i} to \thirdparty, the simulator instead encrypts arbitrary messages; this is indistinguishable from the honest messages of \Ppay because of the IND-CPA security of the encryption scheme.\\

\subsubsection*{Sub-protocols of Sub2 - Labeled D-PSI}
\noindent We now present three sub-protocols of the labeled D-PSI:
\paragraph{sub2.1 - D-PSI} This is the application of the $(n,n)$-threshold encryption to the identifiers. To obtain the intersection of the identifiers, the data providers encrypt the identifiers $\id i k$ with $\tilde sk^i_k$ and share them with the data collector who can decrypt them if and only if they are part of the intersection.

\paragraph{sub2.2 -Labeled D-PSI with payload} In this protocol, data from different providers is linked in a privacy-preserving
manner. In this case, the input sets are extended with attributes held by each data provider. Again using the $(n,n)$-threshold encryption applied to each record in the local data set, the ciphertext is passed to the data collector, who can decrypt the records corresponding to identifiers that are part of the intersection. The identifiers themselves, are not revealed.

\paragraph{sub2.3 - Threshold Labeled D-PSI with payload}
Labeled D-PSI with payload can easily be extended with an additional {\bf $(t,m)$-threshold encryption}, such that the payload can only be decrypted when the size of the intersection exceeds some threshold $t$.
To that end, each data provider \party i chooses a random secret $\shsecret i \ $ and splits it in $m=| X^i |$ shares $\shshare i k$ using Shamir's $(t, m)$-secret sharing scheme:
\begin{align*}
  \shsecret i \                     & \xleftarrow[]{\$} \{0,1\}^{\kappa}                \\
  \{\shshare i 1, ..,\shshare i m\} & \leftarrow \tsssplit{t, m,\kappa}(\shsecret i \ )
\end{align*}
%\jl{Check with Cyprien + key re-use: $\tilde {sk}_k^i$ is used both to encrypt the payload as a one time pad, and as the $(n, n)$-threshold encryption key}
%\cdsg{I think this is fine because the collector doesn't have any information about $\tilde{sk}_k^i$ for the entries outside of the intersection other than one ciphertext.}
Each payload record is extended with one unique share $\shshare i k$ which is encrypted with the $(n,n)$-threshold encryption key $\tilde sk^i_k$ (see above) while the payload is encrypted using ($\shsecret i \  \oplus \tilde sk^i_k$). The message each data provider \party i sends to the data collector \thirdparty now includes:
\begin{align*}
  \Bigl\{\bigl(\ldots, \enc_\kappa(\shsecret i\oplus \tilde sk^i_k,data^i_k), \enc_\kappa(\tilde sk^i_k,\shshare i k)\bigr) \Big| k\in[m]\Bigr\}
\end{align*}
If the intersection size is at least $t$, the data collector will be able to decrypt enough shares $\shshare i k$ to reconstruct the random secret $\shsecret i \ $, and hence, compute the decryption key of the payload itself. Otherwise, the payload remains hidden.
Note that the size of the intersection is always revealed.

Different data providers may use different thresholds $t^i$. In that case, some of the payload data may remain hidden (when the intersection size is smaller than $t^i$ for some \party i), while other parts can be decrypted.\\

% \paragraph{\em[sub2.4] k-anonymity PRL.}
%%%%%%%%%% This gives not enough info about k-anonymity and whether the sketched solution is viable, or would provide k-anonymity. e.g., will it provide all or nothing, or only prevent that specific set of records to be decrypted; may not be possible using the core protocol (as it will reveal the number of suppressed records) %%%%%%%%%%

% This scheme can further be extended to realize {\bf \textit{k}-anonymity} towards the third party~\cite{S02}. A provider \party i may denote one or more fields in the payload as being privacy-sensitive. As $k$-anonymity on the input sets may no longer hold once they are linked with the input sets of the other input parties, $k$-anonymity needs to be applied on the intersection.
% We briefly sketch a possible output protocol that supports $k$-anonymity applied to the intersection.

% Let $w$ be the number of unique (combinations of) values of these fields. Provider \party i chooses $w$ random secrets $\shsecret i v$ and splits each in $|X_i|$
% shares. Each payload is then extended with an {\it appropriate share} so that the secret can only be recovered when at least $k$ elements with the same value (combination) belong to the intersection.

The output protocols above demonstrate how the SIKA protocol can be leveraged to permit privacy-enhancing protocols.
Moreover, combinations of the output protocols are also possible. Although more efficient solutions for both {\it sub1} and {\it sub2.1} may exist, these examples illustrate the flexibility of our approach.

It is also worth noting that while the underlying SIKA functionality, presented in Section~\ref{sct:sika}, assumes that the input sets \ID i have equal size $m$ across all parties, this assumption often does not hold in practice. However, this can be easily mitigated by padding smaller input sets with dummy records. Alternatively, to reduce communication overhead, input parties with smaller input sets may encode smaller OKVSs. However, this may leak information about the input set sizes to other entities in the protocol.
\section{LetheLink: Linking data from multiple data providers with unique identifiers}\label{sct:lethelink}

This Section introduces \textsf{LetheLink}, a practical implementation for linking data based on the requirements outlined in Section~\ref{sec:req}.

\textsf{LetheLink} is a command-line interface (CLI) tool that implements the SIKA protocol and two of the payload sub-protocols: Labeled D-PSI with payload ({\it sub 2.2}), and its threshold variant ({\it sub 2.3}). It is designed to demonstrate the protocols' practical applicability. \textsf{LetheLink} integrates with the controller's existing architecture, and supports cleartext analysis of the data. Furthermore, no information is leaked to the input parties, and the amount of information revealed to the collector is strictly minimized. Personal identifiers are no longer exposed, and data is only available to the collector if it is part of the intersection. In fact, sensitive data is only revealed if the intersection is sufficiently large. This integration is essential for our governmental partner to improve privacy and trust, improving step~3 from current practices.

\paragraph{Procedure}
Before the actual protocol is executed, a scientific board prepares and digitally signs a JSON configuration file, which is distributed to the input parties and collector.
%that will forward it to the involved data providers.
%
This file contains all the parameters and connection information required to execute the protocol. Additionally, it contains a human-readable description of the required data, based on which each input party prepares a comma-separated data file (i.e., a CSV file) containing the potentially relevant (identified) personal data. 
To execute the protocol, input parties and the collector use the same CLI client application, providing the shared JSON file. Additionally, each input party provides its own CSV file. Upon successful completion, the collector's client outputs a CSV file containing the combined attributes from all data providers for the records in the intersection.

The remainder of this Section discusses the implementation details and performance results of the SIKA protocol (without payload), and the Labeled D-PSI with payload sub-protocols ({\it sub 2.2} and {\it 2.3}). %Our performance evaluation demonstrates that \textsf{LetheLink} is suitable for the targeted use case involving a limited number of input parties providing data sets covering up to $2^{24}$ individuals.

\subsection{Configuration, Infrastructure and Method}

\begin{table*}[t]
\caption{Total runtime in seconds of the SIKA protocol, starting with loading CSV files by the data providers and finishing after the collector has stored the output CSV file. The intersection size is $m/{2^4}$.} 
\label{tbl:perf-exact}
% \small
% \begin{center}
\begin{tabular}{l||r|r|r|r|r|r || r|r|r|r|r|r ||}
% & \multicolumn{12}{c||}{ \textbf{Computations}} \\
 %\cline{2-13}
 & \multicolumn{6}{c||}{\textbf{LAN}} & \multicolumn{6}{c||}{\textbf{WAN}}\\
\cline{2-13}

 & \multicolumn{3}{c|}{$\kappa = 128, \lambda = 40$} & \multicolumn{3}{c||}{$\kappa = 256, \lambda = 80$} & \multicolumn{3}{c|}{$\kappa = 128, \lambda = 40$} & \multicolumn{3}{c||}{$\kappa = 256, \lambda = 80$}\\
\cline{2-13}

$m$ & $n = 3$  & $n = 5$ & $n = 7$ & 
$n = 3$ & $n = 5$ & $n = 7$ 
& $n = 3$  & $n = 5$ & $n = 7$ & 
$n = 3$ & $n = 5$ & $n = 7$ 
\\

\hline

%\textbf{$2^{16}$} & 3 &  6 & 10 & 3 & 6 & 10 & 4 & 7 & 12 &  5  & 8 & 13\\
%\textbf{$2^{18}$} & 11 & 21 & 37 & 11 & 21 & 35 & 12  & 23 & 38 & 15 & 26 & 45\\
%\textbf{$2^{20}$} & 45  & 86 & 156 & 47 & 88 & 156 & 50 & 91 & 170 & 58 & 108 & 176 \\
%\textbf{$2^{22}$} & 205 & 390 & 667 & 212 & 387 & 667 & 215  & 406 & 688 & 244  & 439 & 757 \\
%\textbf{$2^{24}$} & 983 & 1828 & 2926 & 1068 & 1825 & \multicolumn{1}{c||}{/} & 1024  & 1896 & 2979 & 1141 & 2055 & \multicolumn{1}{c||}{/} \\

%\textbf{$2^{16}$} & 0,7 &  &  & 1,3  &  & & 0,6 &  &  &  &  & \\
\textbf{$2^{16}$} & $< 1$ & 1 & 2 & 1  &  1 & 1 & $< 1$ & 1 & 2 & 1  & 2 & 4\\
\textbf{$2^{18}$} & 2  &  4 & 5 & 3  & 5 & 7 & 4 & 7 & 9 & 6 &  10 & 14 \\
\textbf{$2^{20}$} & 10 & 18 & 26 & 12 & 22 & 30 &  14 & 25 & 36 & 25 & 39 & 54\\
\textbf{$2^{22}$} & 58  & 91 & 117 & 67 & 106 & 145 & 74 & 123 & 167 & 102 & 166 & 241\\
\textbf{$2^{24}$} & 312 & 362 & 486 & 356 & 413 & 570 & 386 & 473 & 674 & 500 & 663 & 961\\

\end{tabular}

% \end{center}
\end{table*}

\begin{table}[t]\caption{Measured data transfer in MB. The volumes sent by a data provider to another data provider and to the collector are denoted by $\party i \rightarrow \party j$ and $\party i \rightarrow \thirdparty$ respectively. 
%Hence, the total incoming volume is $(n-1) (\party i \rightarrow \party j)$ for each data provider and $n (\party i \rightarrow \thirdparty)$ for the collector. The overall communication cost is $n (n-1) (\party i \rightarrow \party j) + n(\party i \rightarrow \thirdparty)$.
%Since $\tau^{e}_n = (n-1)\cdot 2^{e-16} = (n-1)\cdot m \cdot  2^{-16}$ grows linearly in $n$, the total volume received by \thirdparty will increase quadratically in $n$. 
}
\label{tbl:transfer-exact}
\begin{center}
\begingroup
\setlength{\tabcolsep}{6pt} % Default value: 6pt
\renewcommand{\arraystretch}{1.3} % Default value: 1
\begin{tabular}{l|c|c | c|c|}
& \multicolumn{2}{c|}{$\kappa = 128, \lambda = 40$} & \multicolumn{2}{c|}{$\kappa = 256, \lambda = 80$}\\
\cline{2-5}
 $m$ &  \multicolumn{1}{c|}{$\party i \rightarrow \party j$}  & \multicolumn{1}{c|}{$\party i \rightarrow \thirdparty$} &
 \multicolumn{1}{c|}{$\party i \rightarrow \party j$}  & \multicolumn{1}{c|}{$\party i \rightarrow \thirdparty$} \\
\hline
%\textbf{$2^{10}$} & $0.05$& $0.08 + \tau^{10}_n$ & $0.11$ & $0.08 + 2 \cdot \tau^{10}_n$\\
%\textbf{$2^{12}$} & $0.19$ & $0.20 + \tau^{12}_n$ & $0.37$& $0.33 + 2 \cdot \tau^{12}_n$ \\
%\textbf{$2^{14}$} & $0.72$ & $0.81 + \tau^{14}_n$& $1.37$ & $1.3 + 2 \cdot \tau^{14}_n$ \\
%\textbf{$2^{16}$} & $2.8$ & $3.2 + \tau^{16}_n$ & $5.3$& $4.1 + 2 \cdot \tau^{16}_n$\\
\textbf{$2^{16}$} & $3$ & $2 + n$\ \ \ & $7$& $3 + 2n$\\
\textbf{$2^{18}$} & $12$ & $9 + 4n$ & $24$ & $13 + 8n$\ \ \ \\
\textbf{$2^{20}$} & $46$ & $37 + 16n$ & $89$ & $53 + 32n$ \\
\textbf{$2^{22}$} & $181$ & $150 + 64n$\ \ \ & $344$ & $214 + 128n$ \\
\textbf{$2^{24}$} & $713$ & $607 + 256n$ & $1352$ & $863 + 512n$ \\
\end{tabular}
\endgroup
\end{center}
\end{table}

The client is built in Java 17. For the OKVS, we selected and integrated the multi-threaded clustering-based C++ implementation of the scheme of Raghuraman et al.~\cite{raghuraman2022blazing, RindalCode} into \textsf{LetheLink} using the Java Native Interface (JNI). The length of the values in the OKVS is set to twice the key length $\kappa$, matching the requirements of the SIKA protocol. 
The pseudo-random permutation with a one-time key, as required by SIKA, is implemented with AES in CBC mode with zero initialization vector. For the pseudorandom number generator (PRNG), we use \texttt{SHA1PRNG}. 

Performance tests were conducted on AWS EC2 r7i.8xlarge VMs, with 32 vCPUs (Intel Xeon Platinum 8588C processors at a frequency of 3.2~Ghz) and 256 GB~RAM. Core utilization is optimized in our multithreaded application to maximize performance.
The machines used storage with 3000 IOPS and a throughput of 125~MB/s. The operating system was Ubuntu 24.04 LTS. 

The LAN and WAN environments simulated using \texttt{tc} Linux command, are configured to reflect typical real-world conditions, with speeds of 1~Gbps and 150~Mbps and latencies of 1~ms and 30~ms, respectively.

The total runtimes in Table~\ref{tbl:perf-exact}, as well as the runtime overhead reported in Tables~\ref{tbl:perf-prl} and \ref{tbl:perf-shamir}, are averaged over 20 runs for $m \le 2^{22}$ and over 10 runs for $m = 2^{24}$. Detailed performance logs are available as supplementary artifacts and back the numbers mentioned in this section. The runtimes exclude starting the \textsf{LetheLink} client and establishing connections.

%The tests also passed for $n=3$ and $m\le2^{22}$ on virtual machines with 32 GB RAM and 4 vCPU. This resulted in a performance penalty of around 25\% and demonstrates the practicality of the protocol on less powerful machines.

%Ubuntu 22.04 virtual machines were located in different geographical areas, namely London, São Paolo, Sydney, Virginia, Frankfurt, Mumbai, Montreal, and California. The total execution time is the total computation time (see table \ref{tbl:perf-exact}) plus the total communication time (see \ref{tbl:transfer-exact}).  %I/O operations are excluded from the measurements.

%for the input parties and the third party separately. Input parties do all their computations in parallel. Only when this is completed, $\thirdparty$ can do her part. Therefore, the total computation time of the SIKA protocol is the computation time required by the slowest $\party i$ plus the computation time required by $\thirdparty$. 

%For data sources, this comprises the generation of randomness, OKVS encoding and decoding, and the calculation of the $\bosecret j k$ values. For the third party, this is the computation of the $\osecret i k$ values and the computation of the intersection. 

\subsection{SIKA Protocol}
In this Section, we evaluate the performance of the SIKA Protocol, which returns only secrets and pseudonyms returned to the input parties, while the collector obtains those for the intersection. Although SIKA alone is not sufficient as a complete application, this enables us to assess the core performance and scalability of \textsf{LetheLink}.
Table~\ref{tbl:perf-exact} and ~\ref{tbl:transfer-exact} show the total runtime and data transfer of the SIKA protocol under varying conditions. These results provide insights into the efficiency and scalability of the protocol in different network environments.

%The LAN and WAN were configured with speeds of 1 Gbps and 150 Mbps and latencies of 1ms and 30ms respectively, to reflect typical real-world conditions. The scenarios were simulated using \texttt{tc} Linux command, which enables precise control over network characteristics.
%
The computational and statistical security are $\kappa$ and $\lambda$, respectively. The OKVS operations were evaluated with two combined security levels: $(\kappa=128, \lambda=40)$, and $(\kappa=256, \lambda=80)$\footnote{Due to implementation restrictions of the OKVS by Raghuraman et al., the current implementation is limited to only handle statistical security up to 64. Overriding this limit is possible, but results in performance results that are slightly more positive than they should be if $\lambda = 80$.}. Furthermore, the weight parameter, specific for this type of OKVS, was set to the recommended value 3 and hardware acceleration enabled.

\paragraph{Computation}
 %The performance tests include {\it I/O operations, random number generation, computations, and communication times}.
As shown in Table~\ref{tbl:perf-exact}, the runtime increases significantly with larger values of $m$, primarily due to higher computational and communication demands. For example, in the LAN setting with $\kappa=128$, the runtime increases from less than one second when $m=2^{16}$ to more than five minutes when $m=2^{24}$ (for three data providers). The protocol behaves slightly superlinear in $m$; increasing $m$ by a factor of four, as done in the experiments, results in an increase in the total runtime of up to factor 5.8 and by a factor 4.4 on average. 

Due to concurrency between and inside clients, the experiments demonstrate a sublinear behavior of the total runtime with respect to the number of data providers, provided sufficiently fast connections. For example, switching from three to seven data providers results in an increase of less than 60\% of the overall execution time for $n = 2^{24}$ in the LAN setting. 

Switching from the LAN to WAN setting evidently impacts the total runtime. For instance, for $m=2^{24}$, we observe an increase by 24\% for $n=3$, $\kappa=128$, and $\lambda=40$ up to 69\% for $n=7$, $\kappa=256$, and $\lambda=80$. Data processing and communication occur concurrently, but the latter becomes more dominant over slower connections.   

The runtimes summarized in Table~\ref{tbl:perf-exact}
become more stable -- and hence, predictable -- for larger $m$.  The highest relative standard deviation (RSD) decreases from 8.6\% for $m=2^{20}$ to 4.8\% for both $m=2^{22}$ and $m=2^{24}$. Also, more communication overhead results in more stable runtimes; by switching from LAN to WAN, the average RSD decreases from 6.6\%, 2.3\% and 3.4\% to 1.8\%, 2.2\%, and 1.0\%, for the aforementioned values of $m$ respectively.   

%However, thanks to a high degree of parallelism incorporated in the protocol, we can effectively mitigate the time increase despite the substantial growth in both processing and communication workloads.

The computations are dominated by the pseudonym unblinding at the collector; for $n=2^{24}$ this takes between 31\% and 51\% of the total runtime in the LAN setting. The OKVS encode and decode operations, performed $n-1$ times by each data provider, contribute only marginally to the total runtime. For example, encoding $2^{24}$ items takes approximately nine seconds, while decoding requires about five seconds ($\lambda=40; \kappa=128$). 

The total runtime also includes several additional tasks. For the data provider, this encompasses, among others, reading the input file, preparing inputs for the OKVS and creating and sending the messages. For the collector, this encompasses, among others, receiving and parsing the messages, calculating the intersection and writing the output files. 

The intersection size only marginally impacts the total runtime; for $m = 2^{24}$, calculating the intersection and storing the output took maximum 6.4\% and 0.5\% of the total runtime respectively. 

\paragraph{Data transfer} Table~\ref{tbl:transfer-exact} presents the measured data transferred (in MB for $n=3$). 
Each data provider sends data to the $n-1$ other providers ($\party i \rightarrow \party j$), consisting primarily of the OKVS, scaling linearly with the number of records.
The volume sent from a provider to the collector ($\party i \rightarrow \thirdparty$) includes pseudonyms paired with corresponding $z$-values. 
In Table~\ref{tbl:transfer-exact}, we distinguish between a fixed-size portion and a variable portion that grows with increasing number of parties. The fixed part includes the pseudonym and overhead from the current encoding, while the variable part can be estimated as $n\cdot m \cdot  \kappa$ (in bits), representing the volume of the $z$-values introduced per data source.

The total communication cost is thus $n (n-1)$ times the volume sent from a data provider \party i to a data provider \party j plus $n$ times the volume sent from \party i to \thirdparty. For $m = 2^{24}$, this corresponds to 8.2 GB and 15.0 GB for $\kappa = 128$ and $\kappa = 256$, respectively.
Consequently, the total data volume received by \thirdparty increases quadratically in $n$, although this is not reflected in the runtimes of the conducted experiments.

\subsection{Labeled D-PSI with payload}

\begin{table}[t]
\caption{Runtime overhead in seconds of the Labeled D-PSI (sub2.2) extension for $n=3$, with $l$ the byte size of the record payload. }
\label{tbl:perf-prl}
% \small
% \begin{center}
\begin{tabular}{l||r|r|r|r||r|r|r|r||}
& \multicolumn{4}{c||}{\textbf{LAN}} & \multicolumn{4}{c||}{\textbf{WAN}} \\
\cline{2-9}
& \multicolumn{2}{c|}{$\kappa = 128$} & \multicolumn{2}{c||}{$\kappa = 256$} & \multicolumn{2}{c|}{$\kappa = 128$} & \multicolumn{2}{c||}{$\kappa = 256$}\\
\cline{2-9}
 $m$ \textbackslash ~$l$ & $2^6$ & $2^8$ & $2^6$ & $2^8$ & $2^6$ & $2^8$ & $2^6$ & $2^8$ \\
 \hline
 $2^{16}$ & < 1 & < 1 & < 1 & < 1 & < 1 & 1 & < 1 & 1\\
$2^{18}$ & < 1  & 2 & < 1 & 2 & 1 & 5 & 2 & 5 \\
$2^{20}$ & 3 & 7 & 5 & 8 & 7 & 20 & 7 & 20 \\
$2^{22}$ & 12 & 30 & 14 & 34 & 17 & 71 & 29 & 83 \\

%$2^{16}$ & 1 & 1 & 6 & 5& 1 & 1 & 5 & 5\\
%$2^{18}$ & 3 &4 & 17 & 18&4 &5 & 19 & 20 \\
%$2^{20}$ & 15 & 18 & 61 & 67 &11 & 22 & 68 & 74\\
%$2^{22}$ & 78 & 82 & 276 & 268 &78 & 105 &271& 293\\
\end{tabular}
% \end{center}
\end{table}

Table \ref{tbl:perf-prl} presents the runtime overhead of the Labeled D-PSI with payload extension ({\em sub2.2}) for $n=3$. The relative standard deviation of the total runtimes of the experiments ranges from 0.8\% to 8.3\% for $n \ge 18$. 

To keep memory use within limits, the collector stores received messages before processing them. This goes at the expense of time efficiency.
In the LAN setting, I/O operations account for 40\% to 100\% of the additional time for $m \ge 2^{18}$. In contrast, in the WAN setting, communication once again becomes the dominant factor, contributing between 59\% and 100\% of the additional time.

%Aside from increased transfer times, the overhead is due to I/O operations by the collector; To keep memory use within limits, received messages are first stored and then loaded again. TODO: NUMBERS!!!

%In the WAN setting, the extra transfer volumes

%Data providers have $l = 2^6$ or $l = 2^8$ bytes of data per record. Besides encryption, decryption and data transfer, the runtime also includes I/O operations. This keeps memory use within limits, at the expense of time efficiency.
%storage of the received messages by the collector upon receipt, and loading them again just before potential decryption. This eases memory requirements. 
%
The time complexity of the extension is $O(nm)$, since the collector decrypts the attributes of the $m$ records for each of the $n$ data providers.

%
%Stream ciphers are fast; for payloads of size $2^{10}$ bytes, $2^{20}$  encryptions or decryption require 3.6, seconds on AWS t2.2xlarge VMs. For Shamir secret shares with $\kappa = 128$ and $\kappa=256$, this is 0.5 and 0.6 seconds. Since the ciphertexts size equals the plaintext size, the communication overhead is straightforward.

\subsection{Threshold Labeled D-PSI with payload}

\begin{table}[t]
\caption{Runtime overhead in seconds of the Labeled Threshold D-PSI ({\em sub2.3}) extension for $n=3$ in the LAN setting.}
\label{tbl:perf-shamir}
% \small
% \begin{center}
\begin{tabular}{l|r|r|r|r|r|r|}
& \multicolumn{3}{c|}{$\kappa = 128$} & \multicolumn{3}{c|}{$\kappa = 256$}\\
\cline{2-7}
$m$ \textbackslash ~$t$  & $2^{10}$ & $2^{12}$ & $2^{14}$ & $2^{10}$ & $2^{12}$ & $2^{14}$\\
\hline
%$2^{16}$ & 3 & 8& 41 & 4 & 18 & 80\\
%$2^{18}$ & 5 & 35 & 156 & 13 & 72 & 317\\
%$2^{20}$ & 21 & 137 & 624 & 47 & 282& 1262 \\
%$2^{22}$ & 55 & 528 & 1134 & 187 & 1134 & 

$2^{16}$ & $< 1$ & 2 & 9 & $< 1$ & 3 & 19\\
$2^{18}$ & 2 & 9 & 41 & 4 & 18 & 84\\
$2^{20}$ & 6 & 35 & 164& 16 & 72 & 336\\
$2^{22}$ & 13 & 121 &640 & 53 & 270 & 1317\\
\end{tabular}

% \end{center}
\end{table}

% last cell: 5109

\begin{comment}
\begin{table}[t]
\small
\begin{center}
\begin{tabular}{c||r|r||r|r||}
% & \multicolumn{12}{c||}{ \textbf{Computations}} \\
 %\cline{2-13}
 & \multicolumn{2}{c||}{$\kappa = 128$} & \multicolumn{2}{c||}{$\kappa = 256$}\\
\cline{2-5}
$t$ & $\tsssplit{}$ & $\tssrecon{}$ & $\tsssplit{}$ & $\tssrecon{}$ \\
\hline
\textbf{$2^{10}$} & 37 & 0 & 72 & 0\\
\textbf{$2^{12}$} & 144 & 1 & 287 & 1\\
\textbf{$2^{14}$} & 572 & 10 & 1152 & 21\\
\textbf{$2^{16}$} & 2327 & 159 & 4860 & 352\\
\end{tabular}
\caption{Performance measurements in seconds for Shamir secret sharing with increasing threshold $t$ with $m = 2^{20}$ shares.}
\label{tbl:perf-shamir}
\end{center}
\end{table}
\end{comment}

For the threshold variant ({\em sub2.3}), we used a finite field implementation~\cite{LR} of Shamir secret sharing that enables encoding secrets up to 128
and 192 bits. Larger secrets are split prior to encoding. For instance, a 256 bit secret is split in two 128 bit components. Since Shamir secret sharing offers information-theoretic security, this does not affect security guarantees.

Table \ref{tbl:perf-shamir} shows the runtime overhead of the Threshold Labeled D-PSI in a LAN setting with $n=3$. The computational overhead is primarily incurred by the data providers and is negligible for the collector. As each data provider computes its $m$ shares independently and in parallel to one another, increasing $n$ has a negligible impact on the overall computation time. 
Therefore, for a fixed threshold $t$, the time complexity for the computations approximates $O(m)$.
 
Each share is a $(x,y)$ pair, with $|x| = 32$ and $|y| = \kappa$ bits, resulting in an overhead of 20 and 36 bytes per record sent to the collector for $\kappa = 128$ and $\kappa = 256$, respectively.

The total runtime across experiments exhibited a low RSD, and, hence, a high predictability; ranging from 0.3\% to 3.6\% for $m \ge 2^{18}$.

%Tables~\ref{tbl:perf-prl} and~\ref{tbl:perf-shamir} highlight the feasibility of both protocol extensions, provided the threshold $t$ is chosen carefully. A typical choice is $t = 1000 \approx 2^{10}$.

\subsection{Practical applicability}
Our results demonstrate the practical feasibility of the proposed solution in the targeted use cases, specifically for input set sizes up to $m \le 2^{24}$, small payloads, a threshold $t$ that is typically $1000 \approx 2^{10}$, and rarely more than three data providers. 

The application was developed as a modular and deployable product, with crypto-agility in mind (e.g., enabling easy replacement of the OKVS or PRP implementation), which may affect performance. While the implementation prioritizes maintainability and flexibility over low-level optimization, further code level refinements may still yield additional efficiency gains in terms of computations and memory usage.

The \textsf{LetheLink} implementation supports secure storage of 1) locally generated and calculated secret data, 2) received messages, and 3) state information (i.e., which messages have been scheduled, sent, received, and acknowledged). This feature facilitates crash recovery for individual clients, although it was disabled during the performance evaluation.
Importantly, the secure storage of secret data also enables longitudinal studies, allowing data providers to submit multiple observations over time for the same set of citizens. This can be done without rerunning the entire protocol, thereby improving efficiency and usability in real-world deployments.  

\section{Conclusion \& Future Work}\label{sct:conclusion}

In this paper, we presented a new labeled D-PSI protocol supporting encrypted payload and pseudonymization. A collector without input is the only receiver of the intersection, while the data providers do not learn any information about the the data of other providers, nor the intersection. The core SIKA protocol, a composable building block, is proven secure against semi-honest colluding data providers and a non-colluding semi-honest collector.
It is suitable for the deployment in the existing infrastructure of a trusted party in charge of collecting, linking and pseudonymizing the data, that seeks in reducing its risks and aims at a more privacy-friendly solution.

The source code for the \textsf{LetheLink} application and its protocols, including the possibility to run performance tests, as well as detailed measurements of the conducted experiments, have been made available online~\cite{V20}. The computational performance is practical even for very large data sets and the the network communication amount is limited.

%LetheLink\unsure{add something about our M-PSI protocols?}, a solution for joining distributed data sets providing a practical and more privacy-friendly alternative for current practices. Compared to other proposed solutions, that either require a major reorganization of the infrastructure or a restructuring and/or redistribution of the available data over the entities involved, LetheLink can be deployed in any existing environment where different data sources manage and store their own data sets.
%LetheLink does not require an elaborate setup or pre-installed keys. The data that will be linked is the 'bare' data and does not have to be specially encoded (which can introduce errors, or may hide faults that exist in the data). LetheLink does not leak more information than the data that results from the query. It even allows for additional restrictions where the number of records in the result should exceed a threshold; otherwise, no data will be disclosed.
%The protocol is secure against semi-honest colluding data providers and a non-colluding semi-honest trusted party.

%The computational performance is acceptable even for very large data sets and the amount of data exchanged over the network limited.\jl{practical?}

For {\it future work}, multiple extensions can be created in order to further enhance the usability of our solution.
The protocol can be extended to support {\bf k-anonymity} on attributes in the result. The scientific board may impose {\bf post-processing} steps to be performed by the collector before the data is made available to the researcher. {\bf Selective identification} my be required. For instance, in case the results indicate that a particular citizen has committed fraud, or has a risk to develop a medical condition. {\bf Citizen transparency}, which may be a legal requirement~\cite{P20}, could enable citizens to learn in which projects they are involved.

Finally, the current solution returns the intersection of all input data sets. In some scenarios, however, it is desirable to combine data of multiple sets preserving unmatched records of one or more input sets (i.e., outer joins). The current core protocol prevents this, as only when all parties have a matching identifier linking is possible.

\subsection*{Acknowledgements}
The authors used AI-based tools, including OpenAI ChatGPT, to correct typos, grammatical errors, and awkward phrasing throughout the paper.

%\paragraph{Open Science}
%The source code used for the LetheLink application and its protocols, including the possibility to run performance tests, as well as detailed measurements of the conducted experiments, have been made available online~\cite{V20}.

% \input{ethics}
\bibliographystyle{plain}%Used BibTeX style is unsrt
\bibliography{bibliography}
% \input{appendix}
%\listoftodos
\end{document}